\documentclass[submission, Phys]{SciPost}

\usepackage{orcidlink}
\usepackage{xspace, slashed}
\usepackage{amsmath,amssymb}
\usepackage{graphicx,color}
\usepackage{booktabs}
\usepackage{multirow}
\usepackage{colortbl}
\usepackage[usenames,dvipsnames,svgnames,table]{xcolor}
\definecolor{black}   {RGB}{0.0, 0.13, 0.28}
\definecolor{dukeblue}{rgb}{0.0, 0.0, 0.61}

\usepackage[capitalize]{cleveref}
\usepackage{xfrac}
\usepackage[normalem]{ulem}

\hypersetup{
    colorlinks,
    linkcolor={red!50!black},
    citecolor={blue!50!black},
    urlcolor={blue!80!black}
}
\usepackage{dsfont}

\def\beq{\begin{eqnarray}}
\def\eeq{\end{eqnarray}}

\def\stilde{\widetilde}

\def\cbeta{c_{\beta}}
\def\sbeta{s_{\beta}}
\def\cW{c_{W}}
\def\sW{s_{W}}

\def\fbi{{fb$^{-1}$}}

\newcommand{\smodels}{\mbox{\sc SModelS}\xspace}
\newcommand{\pyhf}{\textsc{Pyhf}\xspace}

\newcommand{\hepdata}{\textsc{HEPData}\xspace}
\newcommand{\hf}{\textsc{HistFactory}\xspace}

\newcommand{\atlasCite}{\cite{ATLAS:2024qmx}\xspace}
\newcommand{\atlasStudy}{ATLAS study~\cite{ATLAS:2024qmx}\xspace}

\newcommand{\rexp} {r_{\rm exp}}
\newcommand{\robs} {r_{\rm obs}}
\newcommand{\rexpComb} {r_{\rm exp}^{\rm comb}}

\newcommand{\mnt}[1]   {m_{\tilde\chi^{0}_{#1} }}
\newcommand{\mch}[1]   {m_{\tilde\chi^{\pm}_{#1} }}

\newcommand{\CH}[3][2]{\Tilde{\chi}_{#3}^{#2}}

\begin{document}

\begin{center}\textbf{
{\LARGE On the coverage of electroweak-inos within the\\[1mm] pMSSM with SModelS}\\[4mm]
{\Large -- a comparison with the ATLAS pMSSM study --}\\[2mm]
}\end{center}

\begin{center}
L\'eo~Constantin\orcidlink{0009-0005-1461-6521}\textsuperscript{1}$^\star$,
Sabine~Kraml\,\orcidlink{0000-0002-2613-7000}\textsuperscript{1},
Andre~Lessa\,\orcidlink{0000-0002-5251-7891}\textsuperscript{2},
Th\'eo~Reymermier\,\orcidlink{0009-0007-5699-180X}\textsuperscript{3},
Wolfgang~Waltenberger\,\orcidlink{0000-0002-6215-7228}\textsuperscript{4,5}
\end{center}

\begin{center}
{\bf 1} Laboratoire de Physique Subatomique et de Cosmologie (LPSC),  Universit\'e Grenoble-Alpes, CNRS/IN2P3, 53 Avenue des Martyrs, F-38026 Grenoble, France
\\
{\bf 2} Centro de Ci\^encias Naturais e Humanas, Universidade Federal do ABC,\\ Santo Andr\'e, 09210-580 SP, Brazil
\\
{\bf 3} Universit\'e de Lyon, Universit\'e Claude Bernard Lyon 1, CNRS/IN2P3, Institut de Physique des 2 Infinis de Lyon, UMR 5822, F-69622, Villeurbanne, France
\\
{\bf 4} Marietta-Blau Institut f\"ur Teilchenphysik,  \"Osterreichische Akademie der Wissenschaften, Dominikanerbastei 16, A-1010 Wien, Austria
\\
{\bf 5} University of Vienna, Faculty of Physics, Boltzmanngasse 5, A-1090 Wien, Austria
\end{center}

\begin{center}
${}^\star$ {\small corresponding author}\\[2mm]
{\small emails: \sf constantin@lpsc.in2p3.fr, sabine.kraml@lpsc.in2p3.fr, andre.lessa@ufabc.edu.br, t.reymermier@ip2i.in2p3.fr, wolfgang.waltenberger@oeaw.ac.at}

\end{center}


\section*{Abstract}
{\bf
 The ATLAS collaboration has recently performed a vast scan of the phenomenological Minimal Supersymmetric Standard Model (pMSSM) with a focus on the electroweak-ino sector, and analysed how their Run~2 searches for electroweak production of supersymmetric (SUSY) particles constrain this dataset. All the SLHA files from the scan as well as the constraints from the eight individual searches considered by ATLAS were made publicly available. We use this material to study how well the ATLAS constraints can be reproduced with \smodels v3.0. Moreover, we explore how the picture changes when also including CMS results, and what can be gained by the statistical combination of analyses. Finally, we discuss the part of parameter space with light electroweak-inos that remains valid despite the stringent LHC limits. Our results underscore the need of a broad, multifaceted approach for maximising sensitivity and closing loopholes in the extensive SUSY parameter space.
}

\vspace{10pt}
\noindent\rule{\textwidth}{1pt}
\tableofcontents\thispagestyle{fancy}
\noindent\rule{\textwidth}{1pt}
\vspace{10pt}

\section{Introduction}  
\label{sec:intro}

The search for new physics Beyond the Standard Model (BSM) has entered an intriguing phase. Despite extensive efforts at the LHC, direct evidence for new physics is yet to manifest, prompting a deeper look at global analyses and re-interpretation of the available data. These re-interpretations allow us to map out a more general constraint overview of the different BSM scenarios, and where potential signatures of new particles could still be hiding. 

A compelling approach to this problem involves the \smodels software \cite{Kraml:2013mwa}, which systematically matches the signatures of complete BSM models to the myriad of constraints published by ATLAS and CMS in the context of simplified models. \smodels~v3.0 incorporates a rich database of cross section upper limits and efficiency maps 
from more than 110 LHC searches for new physics. We have shown in earlier work \cite{Altakach:2023tsd} how a global likelihood approach leveraging multiple searches can significantly increase the sensitivity for complex scenarios, where the BSM signal spreads out in many different final states. 

It is now interesting and relevant to see how the performance of \smodels, which relies purely on public information, compares to a full reinterpretation within the experimental collaborations. The recent ATLAS study on ``ATLAS Run~2 searches for electroweak production of supersymmetric particles interpreted within the pMSSM'' \cite{ATLAS:2024qmx} provides a perfect basis for such a comparison. 
The so-called phenomenological Minimal Supersymmetric Standard Model, pMSSM~\cite{MSSMWorkingGroup:1998fiq}, is a 19-dimensional realisation of the R-parity conserving MSSM with parameters defined at the electroweak scale.  
Its minimal phenomenological assumptions (degeneracy of the first two generations of sfermions, absence of new CP-violating phases, etc.) allow one to capture most of the relevant features of the general MSSM with a quite manageable set of parameters.

In \cite{ATLAS:2024qmx}, the ATLAS collaboration reinterpreted eight separate searches for electroweak SUSY particles, each using 140~fb$^{-1}$ of the LHC Run~2 data taking at $\sqrt{s}=13$~TeV, within the pMSSM. More concretely, they focused on setting constraints on electroweak-inos, i.e.\ charginos and neutralinos, in the pMSSM. The SLHA spectra of the scans performed in \cite{ATLAS:2024qmx} are publicly available on \cite{ATLAS:scandata} together
with information about which point is excluded by which analyses. This is extremely useful
information, which we here use to test the constraining power of \smodels.\footnote{A first such  comparison between ATLAS and \smodels was performed at the time of Run~1 in \cite{Ambrogi:2017lov}.}

In this paper, we compare the constraints on electroweak-inos (EWKinos) derived with \smodels against these new ATLAS pMSSM results. After having laid out our notation and conventions in Section~\ref{sec:notation}, we detail in Section~\ref{sec:setup} the setup of our analysis. Our results are discussed in Section~\ref{sec:results}: First, in Section~\ref{sec:comparisonAtlas}, we perform a direct comparison against Ref.~\cite{ATLAS:2024qmx} by considering only ATLAS EWKino results in the \smodels database and following the same ``single-analysis'' approach as the ATLAS study (that is, we consider the constraint of the most sensitive analysis for each parameter point). In Section~\ref{sec:fulldb} we extend this to the full \smodels database, that is including also CMS results and in particular constraints from gluino production. In Section~\ref{sec:combination}, we then analyse what can be gained by the statistical combination of (approximately) uncorrelated analyses, and we discuss the part of the parameter space with light EWKinos that escapes current LHC limits, paying special attention to small, yet tantalizing excesses. 
A summary and conclusions are provided in Section~\ref{sec:conclusions}.

\section{Electroweak-inos in the (p)MSSM}\label{sec:notation}

The theoretical framework of this paper is the R-parity and CP-conserving MSSM with parameters defined at the electroweak scale. 
The electroweak-ino sector of the MSSM consists of  
four neutralino and two chargino mass eigenstates, denoted as $\tilde\chi^0_{1...4}$ and $\tilde\chi^\pm_{1,2}$, respectively. Our notation and conventions are the same as in \cite{Altakach:2023tsd}; we briefly recall them here for completeness. 
For a discussion of cross sections, decay patterns and resulting LHC signatures, we refer the reader to, e.g., \cite{Altakach:2023tsd,Baum:2023inl}.

Neutralinos are linear combinations of the bino $\stilde B$, neutral wino $\stilde W^0$ and neutral higgsino $\stilde H_u^0$ and $\stilde H_d^0$ gauge eigenstates, while charginos are linear combinations of the charged wino ($\stilde W^+$ and $\stilde W^-$) and charged higgsino ($\stilde H_u^+$ and $\stilde H_d^-$) gauge eigenstates.
The respective mass matrices are given by 
\begin{equation}
    {\cal M}_{N} \,=\,
    \begin{pmatrix}
      M_1 & 0 & - \cbeta\, \sW\, m_Z & \sbeta\, \sW \, m_Z\cr
      0 & M_2 & \cbeta\, \cW\, m_Z & - \sbeta\, \cW\, m_Z \cr
      -\cbeta \,\sW\, m_Z & \cbeta\, \cW\, m_Z & 0 & -\mu \cr
      \sbeta\, \sW\, m_Z & - \sbeta\, \cW \, m_Z& -\mu & 0 \cr
    \end{pmatrix}
\label{neutralinomassmatrix}
\end{equation}
and
\begin{equation}
    {\cal M}_{C} \,=\,
    \begin{pmatrix}
       {\bf 0}&{\bf X}^T\cr
       {\bf X} &{\bf 0}
    \end{pmatrix} \,, \qquad
    {\bf X} \,=\, \begin{pmatrix}
        M_2 & \sqrt{2}\, \sbeta\, m_W\cr
        \sqrt{2}\, \cbeta\, m_W & \mu \cr \end{pmatrix} \,.
\label{charginomassmatrix}
\end{equation}
Here, $M_1$, $M_2$ and $\mu$ are the bino, wino and higgsino mass parameters, respectively, and we have introduced the abbreviations $\sbeta = \sin\beta$, $\cbeta = \cos\beta$, $\sW = \sin\theta_W$, and $\cW = \cos\theta_W$, with $\theta_W$ the weak mixing angle and $\tan\beta=v_2/v_1$ the ratio of the Higgs vacuum expectation values. The chargino mass matrix is written in $2\times 2$ block form for convenience.
The mass eigenstates are related to the gauge eigenstates by the unitary
matrices $N$, $U$ and $V$, which diagonalise the mass matrices:
\begin{equation}
    N^* {\cal M}_{N} N^{-1} = \textrm{diag}\left( \mnt{1}, \ldots ,\, \mnt{4}\right),
    \label{eq:n1_mixing}
\end{equation}
\begin{equation}
    U^* {\bf X} V^{-1} = \textrm{diag}\left( \mch{1},\, \mch{2} \right),
\end{equation}
so that
\begin{equation}
    \tilde\chi^0_i = N_{ij} \tilde{\psi}^0_j \,, \qquad
    { \tilde\chi^+_1 \choose \tilde\chi^+_2 } = V\, {-i\stilde W^+ \choose \stilde H_u^+} \,, \quad
    { \tilde\chi^-_1 \choose \tilde\chi^-_2 } = U\, {-i\stilde W^- \choose \stilde H_d^-} \,,
    \label{eq:c1_mixing}
\end{equation}
where $\tilde{\psi}^0 = \left( -i \tilde B, -i \tilde W^0, \tilde H_d^0, \tilde H_u^0 \right)^T$.
By convention, the physical states are mass ordered: $|\mnt{i}|<|\mnt{j}|$ for $i<j$ and $\mch{1} < \mch{2}$. Complying with the SUSY Les Houches Accord (SLHA)~\cite{Skands:2003cj}, 
we take the mixing matrices to be real, allowing for {\it signed} (negative) mass eigenvalues for the neutralinos.\footnote{The physical masses are of course always positive.} The lightest neutralino, $\tilde\chi^0_1$, is typically also the Lightest Supersymmetric Particle (LSP) and the dark matter candidate.

The bino, wino and higgsino contents of neutralino $\tilde\chi^0_i$ are given by $|N_{i1}|^2$, $|N_{i2}|^2$ and $|N_{i3}|^2+|N_{i4}|^2$, respectively. Likewise, the wino and higgsino admixtures of chargino $\tilde\chi^+_i$ are given by $|V_{i1}|^2$ and $|V_{i2}|^2$. 
If the mass parameters $M_1$, $M_2$ and $\mu$ in eqs.~\eqref{neutralinomassmatrix} and~\eqref{charginomassmatrix} are sufficiently different from each other, the mixing is small and one ends up with a bino-like neutralino with mass of about $M_1$, an almost mass-degenerate pair of wino-like chargino/neutralino with mass of about $M_2$, and a triplet of higgsino-like neutralinos and chargino with mass of about $\mu$. 
On the other hand, if two or all three mass parameters are similar in size, one ends up with mixed states and a much more complicated phenomenology. Consequently, the experimental signatures of charginos and neutralinos heavily depend on the pattern of $|M_1|$, $|M_2|$ and $|\mu|$. This makes in particular EWKino mass limits in the pMSSM so different from those in simplified models.  

It is also worth noting here that, while the notion of bino-like, wino-like or higgsino-like is often very helpful for understanding phenomenological features (and will be used a lot in the following),  ``pure'' bino, wino or higgsino states, as typically assumed in simplified models, basically do not exist in realistic MSSM scenarios unless one evokes very large hierarchies between $|M_1|$, $|M_2|$ and $|\mu|$.\footnote{Even then a minimal, albeit often negligible, mixing is induced by the off-diagonal terms proportional to $m_Z$ in the mass matrices. This originates from electroweak symmetry breaking.} 

Finally, a comment is in order concerning long-lived charginos. These can arise in wino-like or higgsino-like LSP scenarios, that is for 
$|M_2|\ll |M_1|,\,|\mu|$ or $|\mu|\ll|M_1|,\,|M_2|$, because of the small, loop-induced mass splitting between the $\tilde\chi^\pm_1$ and $\tilde\chi^0_1$ in these cases.  
In the pure wino limit, with all other sparticles taken heavy, this mass splitting is about 160~MeV~\cite{Ibe:2012sx}. For such a tiny mass splitting, the $\tilde\chi^\pm_1$ decays into $\tilde\chi^0_1+\pi^\pm$ or  $\tilde\chi^0_1 + \ell^\pm\nu_\ell$ ($\ell=e,\mu$) with a proper decay length $c\tau_0$ of a few cm, thus leading to a disappearing-track signature. In the higgsino case, the mass splitting is slightly larger, about 350~MeV, and chargino lifetimes accordingly shorter, of the order of mm~\cite{Ibe:2023dcu}. 
Either way, the chargino decay length never reaches the order of (tens of) meters, so the $\tilde\chi^\pm_1$ does not appear as a heavy stable charged particle (HSCP) in collider searches.  
The other important observation is that precision calculations of both the mass spectrum and the decay rates, as carried out in \cite{Ibe:2023dcu}, are essential for accurate predictions.

\section{Setup of the analysis}\label{sec:setup}

As mentioned in the Introduction, the starting point of our study is the ATLAS electroweak pMSSM analysis, Ref.~\cite{ATLAS:2024qmx}. In that analysis, the ATLAS collaboration performed a vast random scan of the 19-parameter pMSSM, with a focus on the electroweak-ino sector. More concretely, two scans were performed: 
(1) an `EWKino' scan aiming at general electroweak-ino production and allowing for bino-like, wino-like, or higgsino-like LSPs; and (2) a
`BinoDM' scan designed to over-sample bino‐like LSP models with $\mnt{1}\lesssim 500$~GeV and satisfying dark matter relic density requirements. 

We here reuse the `EWKino' scan dataset. The parameter ranges that were sampled over in this scan are given in Table~\ref{tab:ewkino_ranges}.  Specifically, $M_1$, $M_2$ and $\mu$ were varied between $\pm 2$ TeV, while squarks of the first two generations as well as sleptons were decoupled at 10~TeV. The gluino mass parameter was varied between 1--5 TeV, the pseudoscalar Higgs mass between 0--5 TeV, and $\tan\beta$ between 1--60. Finally, stop and sbottom mass parameters and third-generation $A$-terms were varied within a few TeV as to achieve a SM-like light Higgs with mass $m_h=120$--$130$~GeV.  

\begin{table}[t]\centering
\caption{Parameter ranges randomly sampled for the `EWKino' scan in \cite{ATLAS:2024qmx}.} 
\label{tab:ewkino_ranges}
\begin{tabular}{ l r r l } \toprule
Parameter & Min & Max & Note \\ \midrule
$M_{\tilde{L}_1}$ (=$M_{\tilde{L}_2}$) & 10 TeV & 10 TeV & Left-handed slepton (first two gens.) mass \\
$M_{\tilde{e}_1}$ (=$M_{\tilde{e}_2}$) & 10 TeV & 10 TeV & Right-handed slepton (first two gens.) mass \\
$M_{\tilde{L}_3}$                      & 10 TeV & 10 TeV & Left-handed stau doublet mass \\
$M_{\tilde{e}_3}$                      & 10 TeV & 10 TeV & Right-handed stau mass \\ \midrule
$M_{\tilde{Q}_1}$ (=$M_{\tilde{Q}_2}$) & 10 TeV & 10 TeV & Left-handed squark (first two gens.) mass \\
$M_{\tilde{u}_1}$ (=$M_{\tilde{u}_2}$) & 10 TeV & 10 TeV & Right-handed up-type squark (first two gens.) mass \\
$M_{\tilde{d}_1}$ (=$M_{\tilde{d}_2}$) & 10 TeV & 10 TeV & Right-handed down-type squark (first two gens.) mass \\
$M_{\tilde{Q}_3}$                      & 2 TeV  & 5 TeV & Left-handed squark (third gen.) mass \\
$M_{\tilde{u}_3}$                      & 2 TeV  & 5 TeV & Right-handed top squark mass \\
$M_{\tilde{d}_3}$                      & 2 TeV  & 5 TeV & Right-handed bottom squark mass \\ \midrule
$M_1$                                & $-2$ TeV  & 2 TeV & Bino mass parameter \\
$M_2$                                & $-2$ TeV  & 2 TeV & Wino mass parameter \\
$\mu$                                  & $-2$ TeV  & 2 TeV & Bilinear Higgs boson mass parameter \\
$M_3$                                  & 1 TeV  & 5 TeV & Gluino mass parameter \\ \midrule
$A_t$                                & $-8$ TeV  & 8 TeV & Trilinear top coupling \\
$A_b$                                & $-2$ TeV  & 2 TeV & Trilinear bottom coupling \\
$A_{\tau}$                           & $-2$ TeV  & 2 TeV & Trilinear $\tau$-lepton coupling \\
$M_A$                                  & 0 TeV  & 5 TeV & Pseudoscalar Higgs boson mass \\
$\tan\beta$                            & 1 & 60 & Ratio of the Higgs vacuum expectation values \\ \bottomrule
\end{tabular}
\end{table}

The SUSY mass spectrum and decay branching ratios were computed with {\sc SPheno}~\cite{Porod:2003um} version 4.0.5beta. 
Valid points are required to have a neutralino LSP, $m_h\sim 125$~GeV, and satisfy LEP as well as B-physics constraints. No dark matter constraints are applied. From 20K points originally sampled, this yields a dataset of 12\,280 points, dubbed the `EWKino' scan in \cite{ATLAS:2024qmx}. 

From these 12\,280 EWKino scan points provided by ATLAS, we further remove points declared ``filtered'' by ATLAS. This removes points with $\mch{1} > 1.2$~TeV and/or production cross-sections too low to yield at least 10 events in the Run 2 data, 
as well as points which violate constraints from the Higgs sector: ATLAS considered an upper limit on BR($h\to$\,invisible) of 0.107~\cite{ATLAS:2023tkt} and a lower limit on the CP-odd Higgs boson mass of $m_A>480$ GeV~\cite{ATLAS:2019nkf}. Since \smodels does not account for these constraints, it is more suitable to remove those points beforehand. In total, 2419 points are thus filtered. 

Moreover, in the remaining set of 9\,736 points, we noticed scenarios with exceedingly small $\tilde\chi^\pm_1$--$\tilde\chi^0_1$ mass differences, resulting in so long-lived charginos that they give HSCP signatures. As discussed in the previous section, this should not occur in the pMSSM. The origin seems to be a numerical
precision issue in the {\sc SPheno} version (4.0.5beta) used in the ATLAS scan.\footnote{This issue is not present in the {\sc SPheno}~4.0.5 stable release version.} In \cite{ATLAS:2024qmx}, considering only missing-energy ($E_T^{\rm miss}$) and disappearing-track searches, ATLAS treated points with very long-lived charged particles as ``unconstrained''. However, since they constitute erroneous cases, in our study we prefer to remove these points from the dataset. Concretely, we removed 776 points with too low $\tilde\chi^\pm_1$--$\tilde\chi^0_1$ mass splitting: 774 points with $\mch{1}-\mnt{1}$ below the 2-loop mass splitting for pure winos parametrized in \cite{Ibe:2012sx}, plus 2 other points, for which the spectra were also erroneous.  

For the resulting set of valid scan points, we added the relevant production cross sections at $\sqrt{s}=8$ and 13~TeV, computed at next-to-leading order (NLO) with Resummino~\cite{Fuks:2012qx, Fuks:2013vua,Fiaschi:2018hgm,Fiaschi:2023tkq} and the PDF set cteq66 \cite{Nadolsky:2008zw}, to the SLHA files. This was done by means of the \smodels-Resummino interface described in~\cite{Altakach:2023tsd}. 
This stage of the analysis was the most CPU-time consuming part of our study.\footnote{Once the cross sections have been computed, \smodels itself is fast: without analysis combination, the typical run time for a single point is between a few seconds to a few minutes, depending on the set of analyses that constrain the point. Run times of the order of minutes occur in particular when one or more of the relevant analyses use pyhf statistical models for the combination of signal regions.} 
The cross section calculation at 8~TeV failed for 7 points featuring either higgsino-like or wino-like LSPs, because the NLO computation diverged. These points were also removed from the dataset.\\

After all this data preparation, we are left with 8\,953 scan points, for which we want to compare constraints from \smodels to the ``official'' ones from ATLAS. 
To this end, we use \smodels~v3.0.0 with the following settings (only the most relevant entries in the {\tt parameters.ini} file are listed): 
\begin{verbatim}
   [options]
   combineSRs = True ; to combine signal regions when possible
   [particles]
   model = share.models.mssm ; path to the BSM model file.
   promptWidth = 1e-11 ; particles with larger widths [GeV] are considered prompt
   stableWidth = 1e-25 ; particles with smaller widths [GeV] are considered stable 
   [parameters]
   sigmacut = 1e-3 ; minimum cross section value [fb] in SLHA decomposition
   minmassgap = 10. ; minimum mass gap [GeV] for mass compression
   [database]
   path = official ; to use the database pickle file from the official release
\end{verbatim}

\noindent 
Specifically, we run \smodels with a {\tt sigmacut} of 10$^{-3}$ fb; this defines the minimum signal cross section considered in the decomposition into simplified models. We furthermore set the {\tt minmassgap} parameter to 10 GeV to achieve a good coverage of higgsino LSP scenarios~\cite{Altakach:2023tsd}. Also note that we set {\tt combineSRs = True}, so the signal regions of an analysis are combined whenever a statistical model or covariance matrix is available.

The output of \smodels provides $r$-values for each result in the database. These $r$-values are inversely proportional to $\mu_{95}$, defined as the $95\%$ confidence level upper limit on the signal strength of a specific BSM signal ($r\equiv 1/\mu_{95}$). For a given point, we refer to the most sensitive analysis as the one which gives the highest {\it expected} $r$-value, $\rexp$. The point is then excluded if the corresponding {\it observed} $r$-value is at least 1, $\robs\geq 1$. 
Note that this approach is statistically more sound than considering simply the most constraining analysis (the one with the highest $\robs$), and it is also the approach used in the ATLAS study~\cite{ATLAS:2024qmx}. 
When statistically combining analyses, we will determine the set of approximately uncorrelated analyses that gives the highest total sensitivity, i.e.\ the highest $\rexpComb$. This will be explained in more detail in Section~\ref{sec:combination}.

\section{Results} \label{sec:results}

\subsection{Direct comparison with ATLAS results} \label{sec:comparisonAtlas}

We start the discussion of results with a one-to-one comparison between ATLAS~\cite{ATLAS:2024qmx} and \smodels~v3.0. 
Table~\ref{tab:ana_smodels} gives an overview of the analyses considered in \cite{ATLAS:2024qmx} and their implementation status in \smodels. We see that six of the eight analyses considered in \cite{ATLAS:2024qmx} are  available in \smodels. For the remaining two analyses, an earlier version (for 36~fb$^{-1}$ instead of the full Run~2 luminosity data) is implemented in \smodels, because the full-luminosity analyses did not provide sufficient public material for their inclusion. In the following, we will collectively refer to the analyses in Table~\ref{tab:ana_smodels} as the `ATLAS EWKino analyses'.  

\begin{table}[!t]
  \centering
  \caption{Summary of the analyses considered in \cite{ATLAS:2024qmx} and their implementation status in \smodels; in two cases, only an earlier version of the analysis, for 36~fb$^{-1}$ instead of full Run~2 luminosity, is available in the \smodels v3.0.0 database.}
  \label{tab:ana_smodels}
  \hspace*{-6mm}
  \begin{tabular}{lclcc}
    \toprule
    Type   & \qquad &  ATLAS ID & \qquad & In SModelS \\
    \toprule 
    0 lepton (fully hadronic) & & \href{https://atlas.web.cern.ch/Atlas/GROUPS/PHYSICS/PAPERS/SUSY-2018-41/}{SUSY-2018-41}\cite{ATLAS:2021yqv} & & YES \\ 
    \rowcolor{gray!10} 1 lepton + 2 $b$-jets & & \href{https://atlas.web.cern.ch/Atlas/GROUPS/PHYSICS/PAPERS/SUSY-2019-08/}{SUSY-2019-08}\cite{ATLAS:2020pgy} & & YES \\
    2 leptons + 0 jets & & \href{https://atlas.web.cern.ch/Atlas/GROUPS/PHYSICS/PAPERS/SUSY-2018-32/}{SUSY-2018-32}\cite{ATLAS:2019lff} & & YES \\
    \rowcolor{gray!10} 2 leptons + jets & & \href{https://atlas.web.cern.ch/Atlas/GROUPS/PHYSICS/PAPERS/SUSY-2018-05/}{SUSY-2018-05}\cite{ATLAS:2022zwa}  & & YES \\
    3 leptons on-shell  & & \multirow{2}*{\href{https://atlas.web.cern.ch/Atlas/GROUPS/PHYSICS/PAPERS/SUSY-2019-09/}{SUSY-2019-09}\cite{ATLAS:2021moa}} & & \multirow{2}*{YES} \\
    \noalign{\global\rownum=1}3 leptons off-shell & &  & &  \\ 
    \rowcolor{gray!10} 4 leptons & & \href{https://atlas.web.cern.ch/Atlas/GROUPS/PHYSICS/PAPERS/SUSY-2018-02/}{SUSY-2018-02}\cite{ATLAS:2021yyr} & & \href{https://atlas.web.cern.ch/Atlas/GROUPS/PHYSICS/PAPERS/SUSY-2017-03/}{SUSY-2017-03}\cite{ATLAS:2018eui} \\
    Compressed (soft leptons) & & \href{https://atlas.web.cern.ch/Atlas/GROUPS/PHYSICS/PAPERS/SUSY-2018-16/}{SUSY-2018-16}\cite{ATLAS:2019lng} & & YES \\ 
    \rowcolor{gray!10} Disappearing Track & & \href{https://atlas.web.cern.ch/Atlas/GROUPS/PHYSICS/PAPERS/SUSY-2018-19/}{SUSY-2018-19}\cite{ATLAS:2022rme} & & \href{https://atlas.web.cern.ch/Atlas/GROUPS/PHYSICS/PAPERS/SUSY-2016-06/}{SUSY-2016-06}\cite{ATLAS:2017oal}\\ 
\bottomrule
\end{tabular}
\vspace*{2mm}
\end{table}

\begin{figure}[t]
    \centering
    \includegraphics[width=0.7\linewidth]{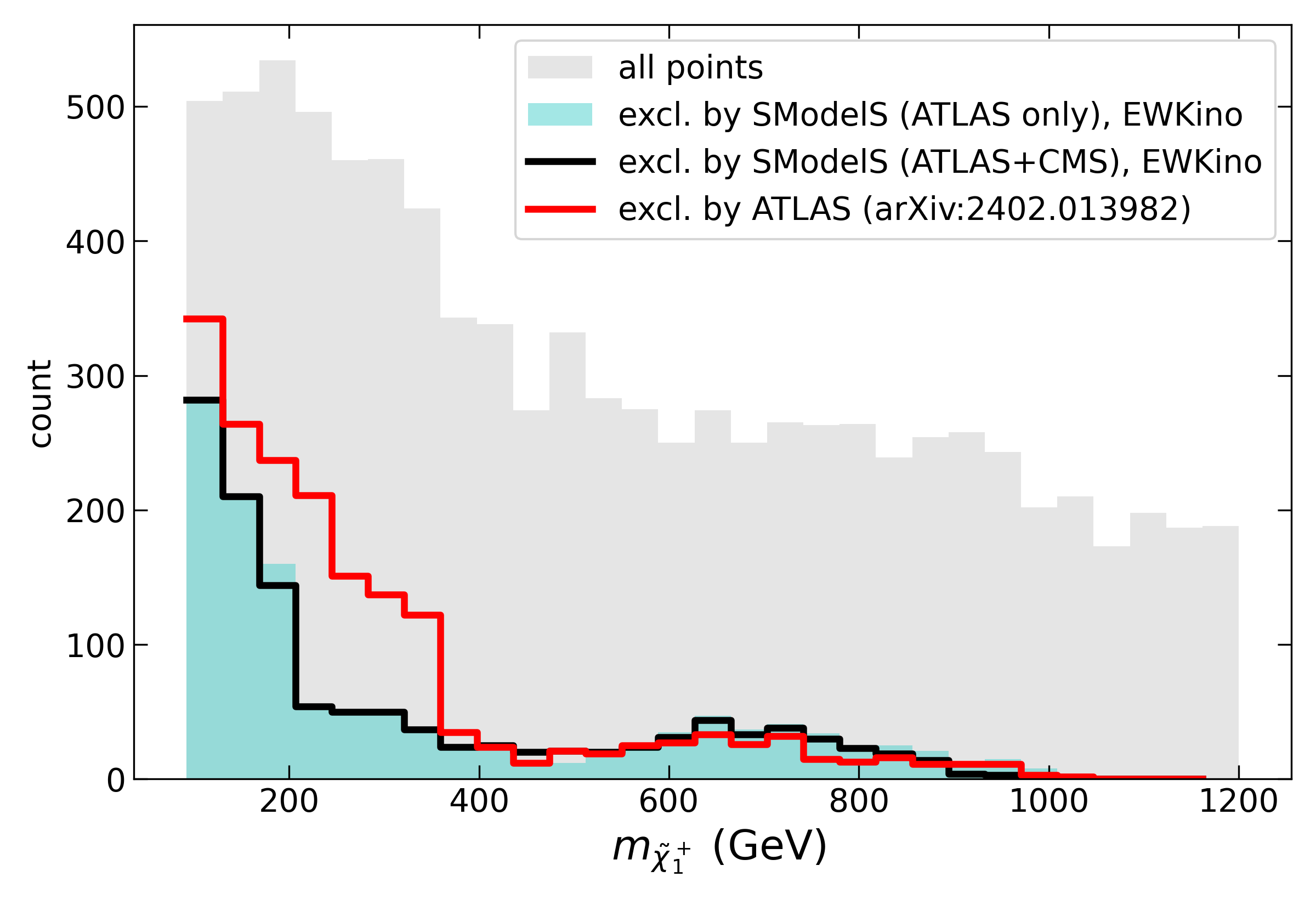}
    \caption{Exclusion by the most sensitive analysis in \smodels, considering only ATLAS EWKino results in the database (cyan) or considering all ATLAS+CMS EWKino results (black). The open red histogram shows the corresponding exclusion from \cite{ATLAS:2024qmx}. The full dataset is represented in grey.}
    \label{fig:histoex}
\end{figure}

Figure~\ref{fig:histoex} compares the number of points excluded by \smodels to that excluded by the \atlasStudy.  
Specifically, the filled cyan histogram shows the number of points excluded by ATLAS EWKino analyses in \smodels as a function of the $\tilde\chi_1^\pm$ mass, while the open red histogram shows the respective official exclusion from ATLAS~\atlasCite. In both cases, a single-analysis approach is taken, meaning that for each scan point only the result of the the most sensitive analysis is considered. 

We generally observe good agreement between ATLAS and \smodels for chargino masses $\mch{1}\gtrsim 360$~GeV, albeit there is a slight over-exclusion from \smodels at high $\mch{1}$, above 600~GeV. This over-exclusion comes primarily from the efficiency map (EM) implementation of the fully hadronic EWKino search ATLAS-SUSY-2018-41~\cite{ATLAS:2021yqv}, as will be discussed in more detail below. 

For $\mch{1}\lesssim 360$~GeV, there is a sizeable difference between the \smodels and the official ATLAS exclusion. This difference arises mainly from the disappearing track constraints: the \smodels v3.0.0 database contains only the 36~fb$^{-1}$ disappearing track results (from ATLAS-SUSY-2016-06~\cite{ATLAS:2017oal}),  yielding weaker constraints on long-lived scenarios compared to the recent 136~fb$^{-1}$ results (ATLAS-SUSY-2018-19~\cite{ATLAS:2022rme}) used in \cite{ATLAS:2024qmx}. Naturally, this affects mostly scan points with wino-like LSP, and some higgsino-like LSP scenarios.   
Additionally, in mixed scenarios, the signal can spread out in various channels, which can be summed up in \smodels only if efficiency maps for the corresponding simplified-model topologies are available for all the signal regions of the relevant analysis. This causes a slight under-exclusion by \smodels for the 3~leptons plus missing transverse energy (MET) search ATLAS-SUSY-2019-09~\cite{ATLAS:2021moa}, as the final constraint would benefit from efficiency maps for the $Wh$+MET or $ZZ$+MET simplified models in addition to the existing ones for $WZ$+MET. In a similar manner, the \smodels implementation of ATLAS-SUSY-2018-05~\cite{ATLAS:2022zwa} would benefit from topologies with Higgs bosons; we will come back to this later in this section. 

So far, we only considered the set of analyses listed in Table~\ref{tab:ana_smodels}. The \smodels database however also comprises equivalent analyses from CMS. (The full list of relevant analyses is given in the appendix.) The exclusion obtained at single-analysis level when considering both ATLAS and CMS EWKino results is shown by the open black histogram in Fig.~\ref{fig:histoex}. We notice that the cyan and black histograms are very similar, indicating a good agreement between the ATLAS and CMS constraints.\\

\begin{figure}[t]
    \centering
    \includegraphics[width=0.49\linewidth]{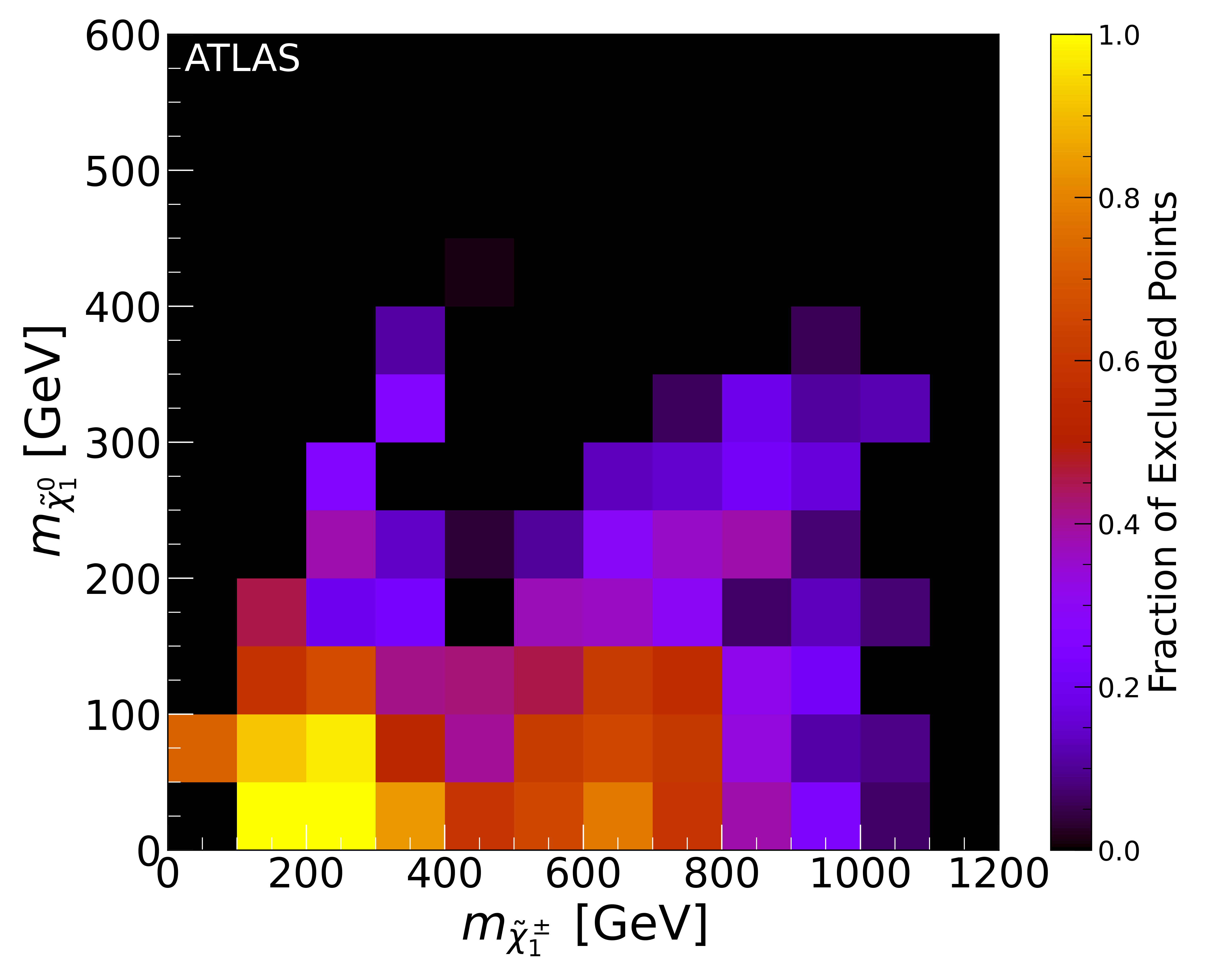}
    \includegraphics[width=0.49\linewidth]{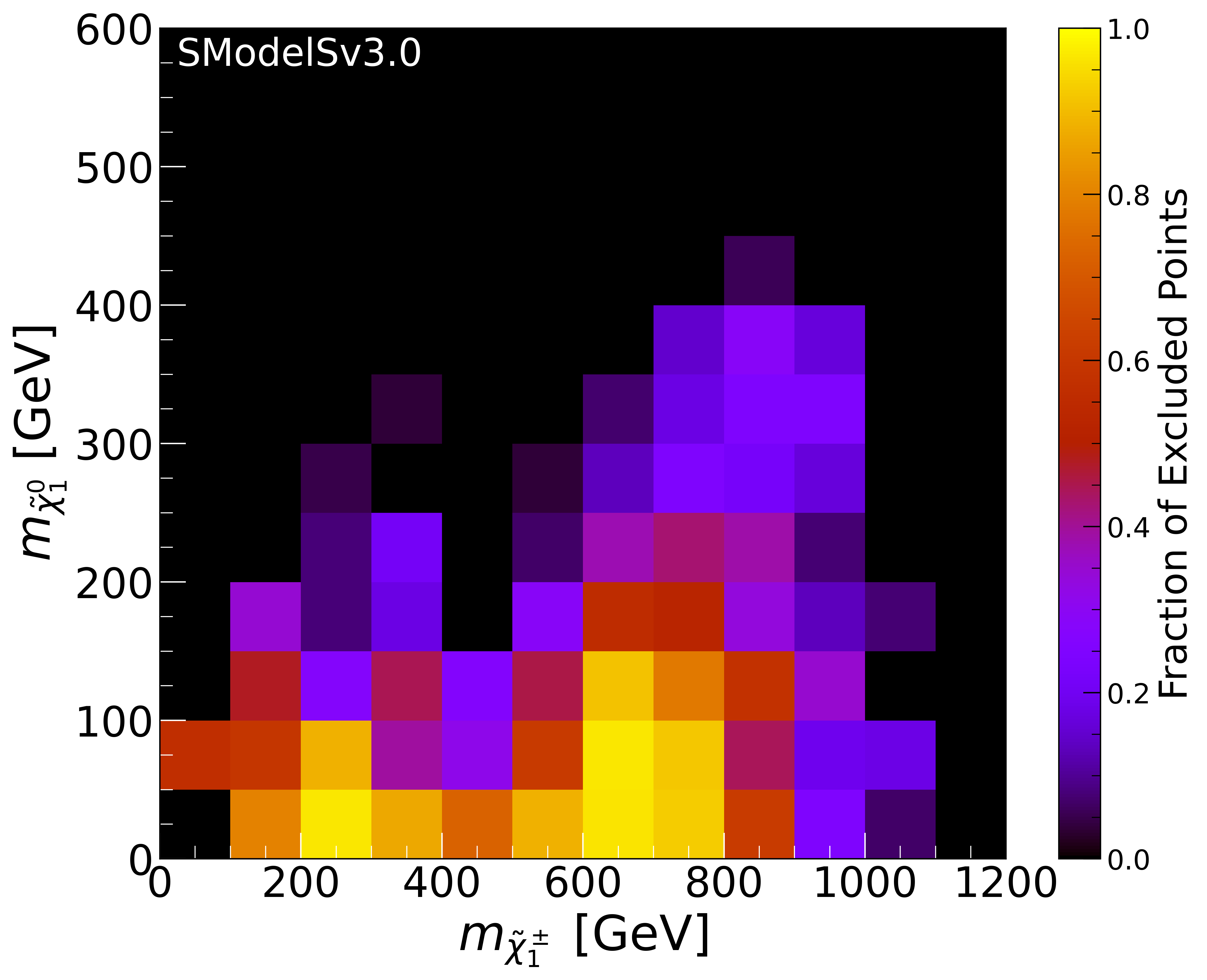}
    \caption{Fraction of points excluded by the ATLAS pMSSM study (left) and by ATLAS EWKino results in \smodels (right). The most sensitive analysis is selected for each point. No points populate the black bins.}
    \label{fig:fracexcl}
\end{figure}

\begin{figure}[!ht]
    \centering
    \includegraphics[width=0.49\linewidth]{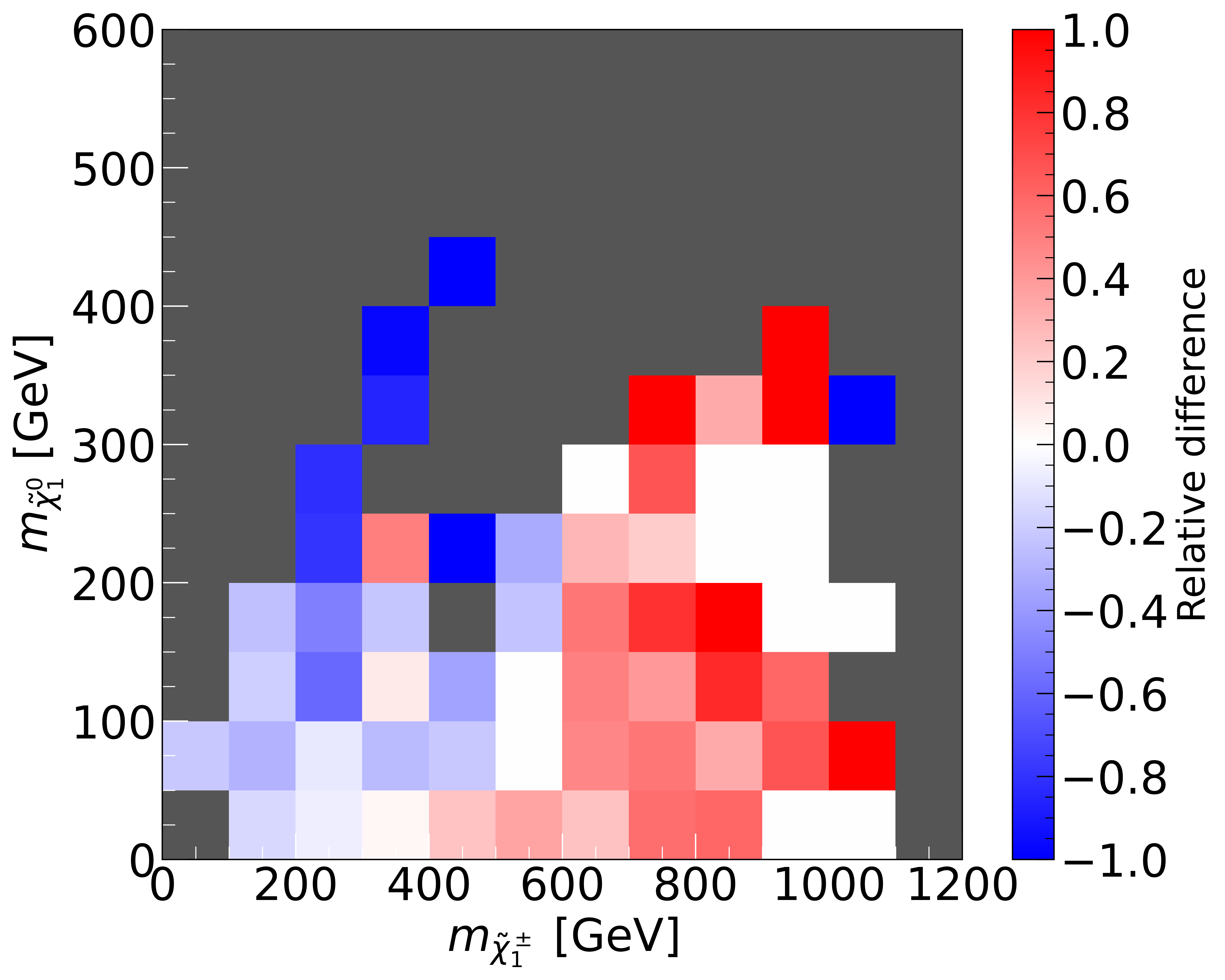}
    \includegraphics[width=0.49\linewidth]{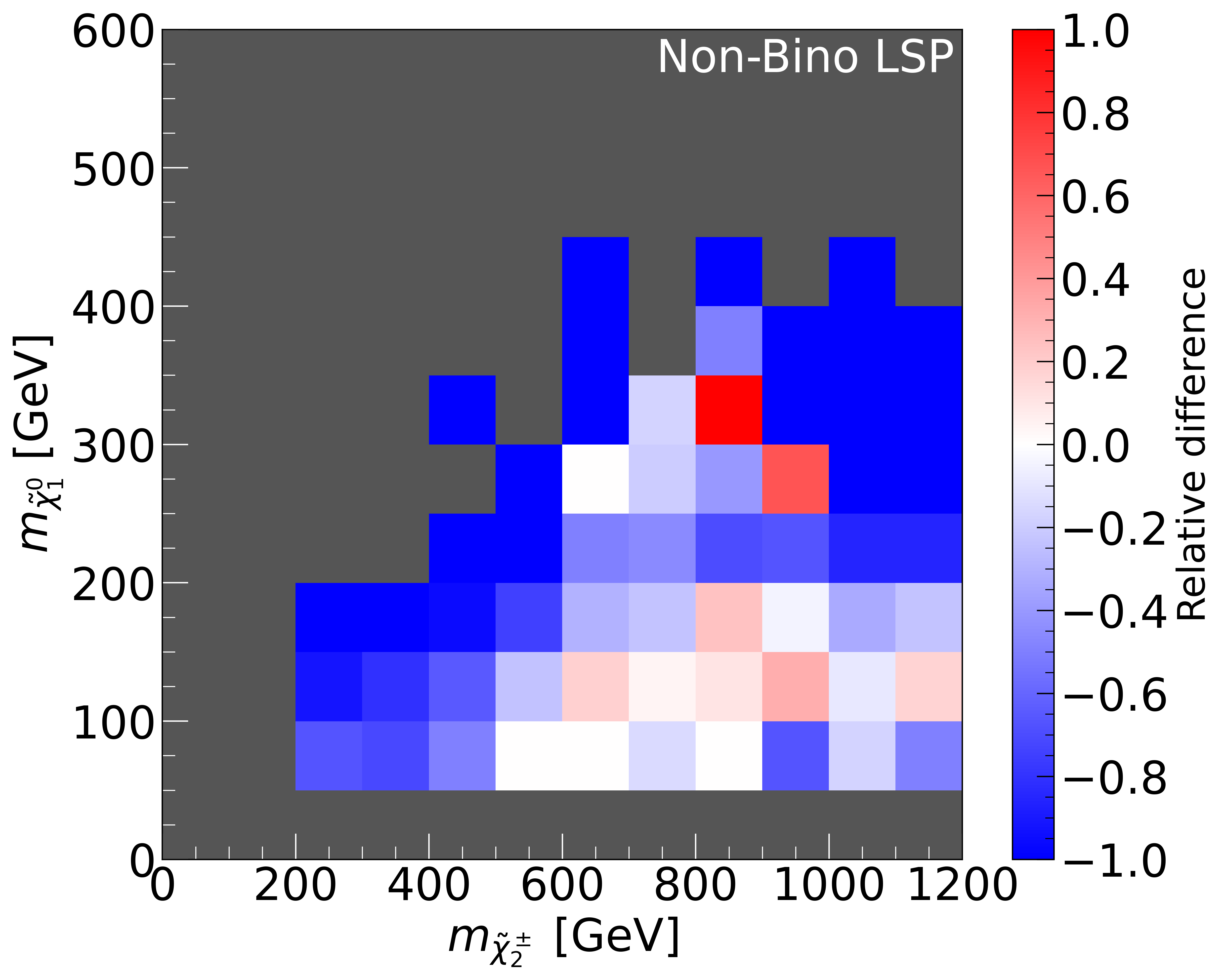}
    \caption{Relative difference between the \smodels and ATLAS exclusions,
    ($\#$excl(\smodels)$-\#$excl(\text{ATLAS}))/$\#$excl(\text{ATLAS}), 
    for the whole scan (left) and for the non-bino LSP points (right). The most sensitive analysis is selected for each point. For \smodels, only ATLAS EWKino results are considered. No points populate the dark grey bins. }
    \label{fig:relativediff}
\end{figure}

To take a closer look at the differences between the \smodels and ATLAS exclusions, it is useful to plot 
the fraction of excluded points as a function of the lightest chargino and the LSP masses, as done in Fig.~\ref{fig:fracexcl}. For most scenarios, the former ($\mch{1}$) controls the overall signal strength, while the latter ($\mnt{1}$) is related to the signal kinematics, such as the hardness of the MET spectrum. 
The ATLAS results from \atlasCite, shown on the left, are especially strong in the low-mass region as the fraction of excluded points by the most sensitive analysis approaches unity for the lower-left corner of the plot. 
The \smodels results using only the ATLAS EWKino analyses 
are shown on the right-hand side: The low-mass region is also well constrained but the fraction of excluded points can also reach (nearly) 1 in the $\mch{1}\approx 600-800$~GeV region. In contrast, the diagonal populated by compressed spectra ($\mch{1}\sim\mnt{1}$) is less constrained in \smodels compared to the \atlasStudy.  Both are in agreement with the previous observations regarding (small) over- and under-exclusions. 

The performance of \smodels is assessed in further detail in Fig.~\ref{fig:relativediff}, where we show the relative difference of the number of excluded points from \smodels and ATLAS. The left panel in Fig.~\ref{fig:relativediff} shows the $\mnt{1}$ vs.\ $\mch{1}$ plane for the complete dataset, while the right panel shows the $\mnt{1}$ vs.\ $\mch{2}$ plane for the subset of non-bino-like LSP points. In both panels, 
blue bins correspond to \smodels excluding fewer points than ATLAS, while red bins correspond to \smodels excluding more. The dark grey bins are empty and indicate mass regions not populated by our dataset or in which no points are excluded.  

Let us first discuss the left panel of Fig.~\ref{fig:relativediff}. 
One can notice a red area in the range $\mch{1}\gtrsim 600$~GeV, corresponding to the previously noted over-exclusion in \smodels. As mentioned before, this comes from the hadronic EWKino search, ATLAS-SUSY-2018-41. Over the whole dataset, we find 222 points excluded by \smodels but not by ATLAS, 166 (75\%) of which are excluded by ATLAS-SUSY-2018-41 in \smodels. The implementation of this analysis is discussed in Appendix~\ref{app:ATLAS search}, with the conclusion that a more conservative version thereof does not resolve the problem; we therefore choose to keep the ATLAS-SUSY-2018-41 implementation in \smodels as it is. In any case, only about 2\% of the whole dataset is affected. Moreover, the issue is mitigated when also considering the hadronic EWKino search from CMS, CMS-SUS-21-002~\cite{CMS:2022sfi}, which is often more sensitive but less constraining than the ATLAS one, cf.\ the black line in Fig.~\ref{fig:histoex}.\footnote{The reason is that CMS-SUS-21-002 observed a small excess while ATLAS-SUSY-2018-41 observed an under-fluctuation of background events; this is discussed on some detail in \cite{MahdiAltakach:2023bdn}.}

In the low mass region, $\mch{1}\lesssim 600$~GeV and $\mnt{1}\lesssim 200$~GeV, good agreement is observed between the results from \smodels and ATLAS. The previously discussed effect of the disappearing track analysis appears on the diagonal: in this region, the LSP is wino- or higgsino-like and the $\tilde\chi^\pm_1$--$\tilde\chi^0_1$ mass difference can become small enough for the chargino to become long-lived. This is especially true for wino-like LSP scenarios (see Section~\ref{sec:notation}). These scenarios are predominantly constrained by disappearing track searches. In Fig.~\ref{fig:relativediff} (both, left and right panels), the different exclusion ranges of the disappearing tracks searches used in the \atlasStudy and in \smodels are clearly visible: only the 136~fb$^{-1}$ result is able to constrain points with LSP and long-lived chargino masses above 200 GeV. (Note here, that the diagonal with $\mch{1}\approx\mnt{1}$ in the left panel corresponds to high $\mch{2}$ in the right panel.)

\begin{figure}[!t]
    \centering
    \includegraphics[width=0.65\linewidth]{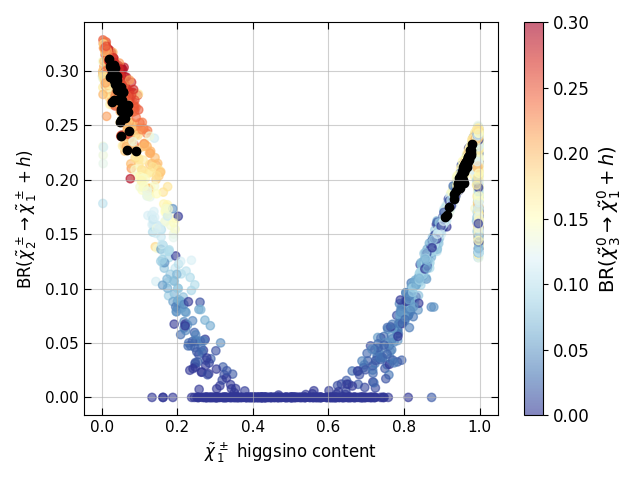}
    \caption{Scatter plot of non-bino LSP points showing BR($\tilde\chi^\pm_2 \to \tilde\chi^\pm_1+h$) versus the $\CH{\pm}{1}$ higgsino content (U$_{12}^2$+V$_{12}^2$)/2. 
    The colour code indicates the branching ratio for $\tilde\chi^0_3 \to \tilde\chi^0_1+h$ decays. The points excluded by ATLAS-SUSY-2018-05 in~\cite{ATLAS:2024qmx} but not excluded by the most sensitive analysis in \smodels are overlaid in black.}
    \label{fig:BRCH2NT3}
\end{figure}

Finally, in the right panel of Fig.~\ref{fig:relativediff}, there is a second region of dark blue bins at roughly $\mch{2}-\mnt{1}\lesssim200$~GeV. 
Points in this region are mostly constrained by the 2~leptons + jets search SUSY-2018-05 in Ref.~\atlasCite, but escape the exclusion with \smodels. As can be seen in Fig.~\ref{fig:BRCH2NT3}, $\tilde\chi^\pm_2 \to \tilde\chi^\pm_1+h$ and $\tilde\chi^0_3 \to \tilde\chi^0_1+h$ decays have sizeable branching ratios in this case. The ATLAS-SUSY-2018-05 implementation in \smodels, however, only constrains the $WZ$+MET part of the signal. Additional efficiency maps for other signal topologies, in particular those involving Higgs bosons, would be helpful for improving the coverage. Furthermore, in the higgsino-like LSP case, a part of the 2~leptons + jets signal comes from $\tilde\chi^\pm_2$ and/or $\tilde\chi^0_{3,4}$ decays to the $\tilde\chi^0_2$, followed by $\tilde\chi^0_2\to \tilde\chi^0_1+ff'$. If $\mnt{2}-\mnt{1}$ is larger than the {\tt minmassgap} parameter, the signal is lost for \smodels (because the $\tilde\chi^0_2\to \tilde\chi^0_1$ decay is not compressed). Again, additional efficiency maps, this time for topologies with an additional step in the decay chain, would be necessary.\\

\begin{table}[!t]
\centering
\caption{Number of points excluded by ATLAS in \atlasCite but not excluded by the most sensitive ATLAS analysis in \smodels; listed per most sensitive analysis of \atlasCite. 
The number after the slash gives the total number of points excluded by the particular analysis in \atlasCite.}
\label{tab:escapingPoints}
\begin{tabular}{llr}
\toprule
ATLAS-SUSY-2018-19 & (disappearing tracks) & 453/871\\
ATLAS-SUSY-2018-05 & (2 leptons + jets) & 124/192\\
ATLAS-SUSY-2019-09 & (3 leptons) & 105/238\\
ATLAS-SUSY-2018-41 & (0 leptons) & 51/386\\
ATLAS-SUSY-2018-16 & (soft leptons) & 19/48\\
ATLAS-SUSY-2019-08 & (1 lept.\ + 2 $b$-jets) & 4/36\\
ATLAS-SUSY-2018-32 & (2 leptons) & 2/27\\
ATLAS-SUSY-2018-02 & (4 leptons) & 2/2\\
\bottomrule
\end{tabular}
\end{table}

\begin{figure}[!t]
    \hspace*{-2mm}\includegraphics[width=0.51\linewidth]{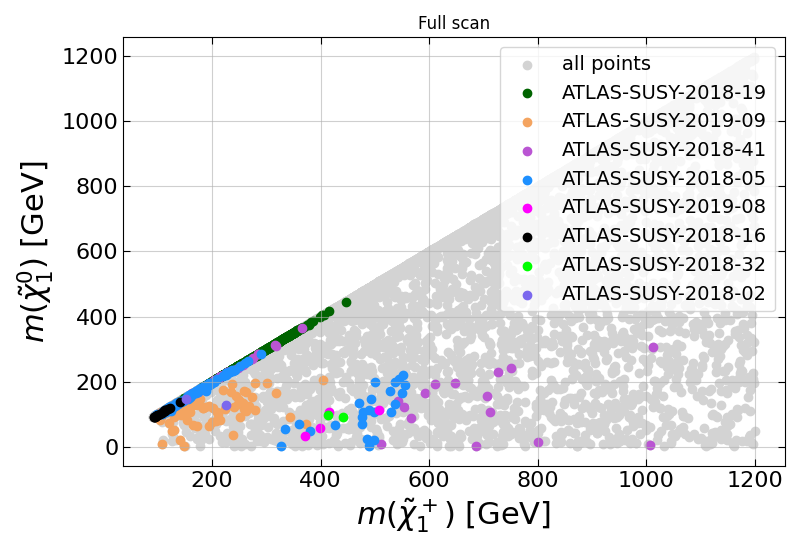}%
    \includegraphics[width=0.51\linewidth]{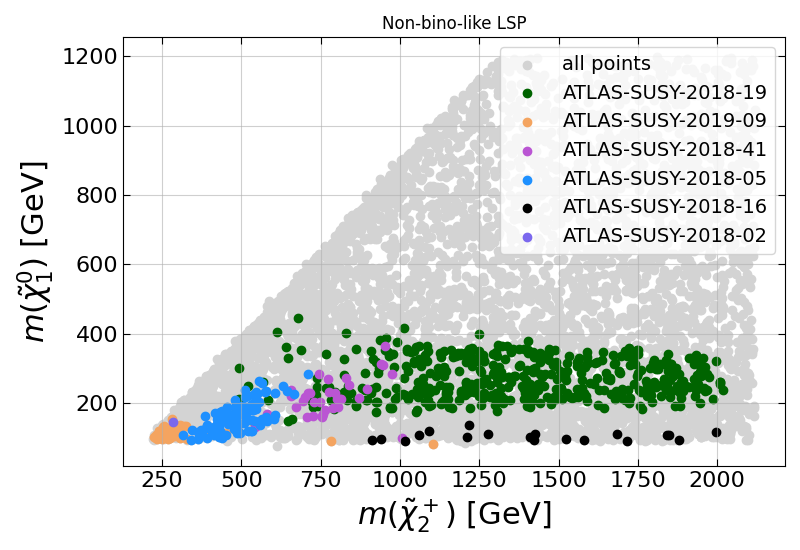}
    \caption{Points (in colour) excluded by ATLAS in \cite{ATLAS:2024qmx}, but not excluded by the most sensitive ATLAS analysis in \smodels, on the left for the full EWKino dataset, on the right for the subset of non-bino-like LSP points. The most sensitive analysis in \cite{ATLAS:2024qmx} is indicated in colour; grey points represent all other scan points, excluded and non-excluded ones.}
    \label{fig:scatterexatlasnonexsmodels}
\end{figure}

To wrap up this section, Table~\ref{tab:escapingPoints} gives the number of points excluded by ATLAS in \atlasCite but not excluded by \smodels (compared to the total number of points excluded by the respective analysis in \atlasCite).\footnote{Note that some scan points have a different most sensitive analysis in \smodels compared to ATLAS; as a result, the $n_{\rm missed}$/$n_{\rm total}$ ratios in Table~\ref{tab:escapingPoints} do  not exactly represent the coverage discussed before.}
In total, 760 points (42\%) are excluded by ATLAS but not by \smodels, 106 with bino-like LSP and 654 with non-bino like LSP. As stated earlier, the largest difference (453 points) comes from the outdated disappearing track results in \smodels. This is followed by the 2~leptons + jets analysis SUSY-2018-05 (124 points) and the 3~leptons analysis SUSY-2019-09 (105 points).
There are also 51 points missed which should be excluded by the fully hadronic search SUSY-2018-41 (as opposed to the 166 points discussed before which are over-excluded by \smodels). 
The distribution of these points in the parameter space is shown in Fig.~\ref{fig:scatterexatlasnonexsmodels}. 

All in all, these observations point to areas where the \smodels database could be improved, if the relevant material was available from the experiments; this concerns in particular the recent disappearing track analyses and efficiency maps for signal topologies involving $h$ bosons, which are relevant for spectra  with higgsino-like LSPs. Indeed this underlines one complication of the simplified model  approach: the reach of these analyses in \smodels is always limited by the diversity of topologies in their implementation.

\subsection{Using the full \smodels database} \label{sec:fulldb}

In the previous subsection we saw that \smodels constraints are generally in good agreement with the official ones from ATLAS when the relevant analyses are available in the  database, though in some cases there remains room for improvement. Let us now go a step further and consider the coverage of the ATLAS EWKino scan by the full \smodels database. This is interesting for two reasons: First, CMS results can give complementary constraints to the ATLAS ones. Second, while squark masses were set to 10~TeV in the ATLAS scan, the gluino mass parameter $M_3$ was allowed to vary between 1--5~TeV. The EWKino dataset thus includes points for which gluino-pair production and/or gluino-EWKino associated production is within reach. 
Although not considered by ATLAS~\atlasCite, the full \smodels database allows us to investigate the complementarity of searches for electroweak and strong SUSY production.

We therefore added the relevant gluino production cross sections for $\sqrt{s}=8$~TeV (for $m_{\tilde g}<2$~TeV) and 13~TeV (for $m_{\tilde g}<3$~TeV) to the SLHA files. For gluino-pair production, we used the NNLL values given by the LHC SUSY Cross Section Working Group~\cite{lhcsxwg,Beenakker:1996ch,Beenakker:2011fu,Beenakker:2013mva,Beenakker:2014sma,Beenakker:2015rna,Beenakker:2016lwe,Kulesza:2008jb,Kulesza:2009kq}, while for 
gluino-EWKino production we computed the cross sections at NLO with  Resummino~\cite{Fuks:2016vdc}.

In total, we consider 36 Run~2 analyses (20 ATLAS and 16 CMS) and 14 Run~1 analyses (6 ATLAS and 8 CMS) from the \smodels v3.0.0 database, for which both observed and expected $r$-values can be computed.\footnote{Analyses, for which only observed cross section upper limit (UL) results are available, are left out, because their sensitivity cannot be determined.} 
An overview is given in Tables~\ref{tab:analist_13tev} and \ref{tab:analist_8tev} in Appendix~\ref{app:Analyses_used}.  
As before, for each scan point we consider only the constraint from the most sensitive analysis, that is the analysis that gives the highest $\rexp$ value.
It is interesting to note here that the number of 1262 points excluded by ATLAS EWKino results in \smodels decreases to 1201 when adding CMS EWKino results. The reason for this is that sometimes a CMS analysis is more sensitive but, due to statistical fluctuations, excludes less than the ATLAS one. When adding also gluino constraints, the number of excluded points increases to 1696.  

\begin{figure}[t]
    \centering
    \includegraphics[width=0.7\linewidth]{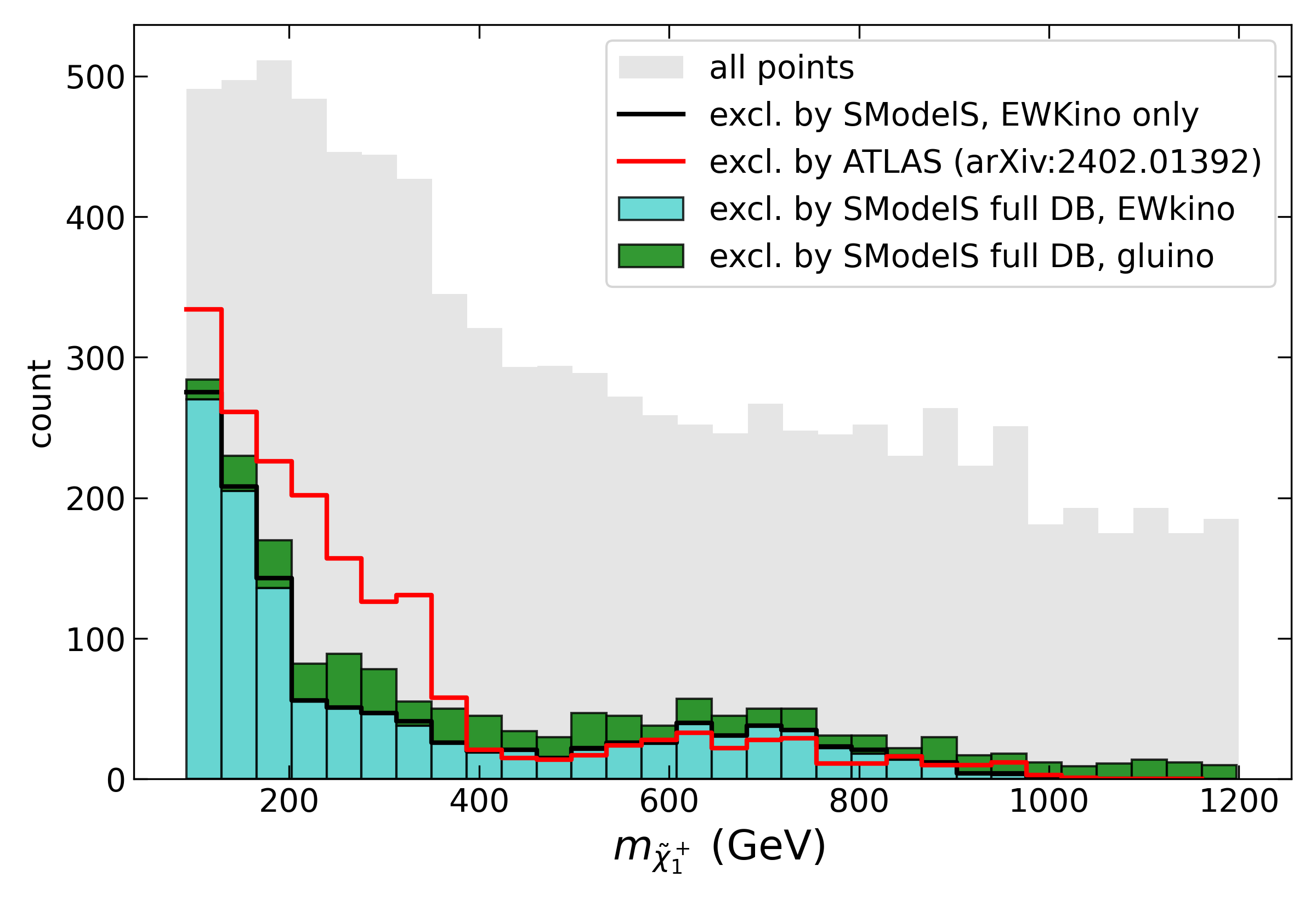}
    \caption{Exclusion by the most sensitive analysis in \smodels, considering only EWKino results in the database (black) or considering all results (stacked histogram). The turquoise and green colors represent the portion of points excluded by EWKino and gluino constraints respectively. The open red histogram shows the corresponding exclusion from \cite{ATLAS:2024qmx}. The full dataset is represented in grey.}
    \label{fig:histoexcompa}
\end{figure}

The exclusion by the most sensitive analysis in SModelS, considering both EWKino and gluino production, is shown in Fig.~\ref{fig:histoexcompa} by the filled stacked histogram, with the turquoise and green colours representing the portions of points excluded by EWKino and gluino constraints, respectively. As can be seen, the gluino constraints are quite relevant over the full range of considered chargino masses, and they allow to access also a part of the parameter space with $\mch{1}\gtrsim 1$~TeV. 

A priori it is possible that a point is excluded by both EWKino and gluino searches. The question then is which type of search is more sensitive, and in particular whether gluino searches generally ``win'' over EWKino ones. An answer to this question can be obtained by comparing the turquoise histogram in Fig.~\ref{fig:histoexcompa} to the black line, which shows the exclusion when considering only EWKino searches. The difference indicates the number of points for which the exclusion by the most sensitive analysis switches from an EWKino to a gluino search. Indeed the difference is very small; it concerns only 30 points, about half of which are in the compressed region with $\mch{1}\lesssim 200$~GeV. All in all, we can conclude that there is little overlap in sensitivity between EWKino and gluino searches in \smodels.\footnote{This may be different in full analysis recasts which, in the presence of a mix of gluino (cascade) decays, may pick up more of the gluino signal than possible with simplified-model results.}

\begin{table}[!t]\centering
\caption{Most sensitive analyses in \smodels (full v3.0.0 database) for the EWKino dataset, including gluino production.}
\label{tab:id_analysis_best}
\begin{tabular}{llrr}
\toprule
Analysis ID & Short description & $\#$pts probed & $\#$pts excl.\\
\midrule
\href{http://cms-results.web.cern.ch/cms-results/public-results/publications/SUS-19-010/index.html}{CMS-SUS-19-010}~\cite{CMS:2021beq} & jets + top- and $W$-tag & 1086 & 345\\
\href{https://atlas.web.cern.ch/Atlas/GROUPS/PHYSICS/PAPERS/SUSY-2018-41/}{ATLAS-SUSY-2018-41}~\cite{ATLAS:2021yqv} & hadronic ($0\ell$) EWK  & 894 & 307\\
\href{https://atlas.web.cern.ch/Atlas/GROUPS/PHYSICS/PAPERS/SUSY-2016-06/}{ATLAS-SUSY-2016-06}~\cite{Aaboud:2017mpt} & disappearing track & 871 & 274\\
\href{http://cms-results.web.cern.ch/cms-results/public-results/publications/SUS-19-006/index.html}{CMS-SUS-19-006}~\cite{Sirunyan:2019ctn} & 0$\ell$ + jets, $\not{\!\!H}_T$ & 778 &201\\
\href{http://cms-results.web.cern.ch/cms-results/public-results/publications/EXO-19-010/}{CMS-EXO-19-010}~\cite{CMS:2020atg} & disappearing track & 241 & 191 \\
\href{http://cms-results.web.cern.ch/cms-results/public-results/publications/SUS-21-002/}{CMS-SUS-21-002}~\cite{CMS:2022sfi} & hadronic ($0\ell$) EWK & 1353 & 148\\
\href{https://atlas.web.cern.ch/Atlas/GROUPS/PHYSICS/PAPERS/SUSY-2019-09/}{ATLAS-SUSY-2019-09}~\cite{ATLAS:2021moa} & 3$\ell$, EWK  & 724 & 122\\
\href{https://atlas.web.cern.ch/Atlas/GROUPS/PHYSICS/PAPERS/SUSY-2018-16/}{ATLAS-SUSY-2018-16}~\cite{ATLAS:2019lng} & 2 soft $\ell$ + jets, EWK &  282  & 39\\
\href{https://atlas.web.cern.ch/Atlas/GROUPS/PHYSICS/PAPERS/SUSY-2018-05/}{ATLAS-SUSY-2018-05}~\cite{ATLAS:2022zwa} & 2$\ell$ + jets, EWK  & 94 & 31 \\
\href{https://atlas.web.cern.ch/Atlas/GROUPS/PHYSICS/PAPERS/SUSY-2018-32/}{ATLAS-SUSY-2018-32}~\cite{Aad:2019vnb} & 2 OS $\ell$  & 125 & 29\\
\href{https://atlas.web.cern.ch/Atlas/GROUPS/PHYSICS/PAPERS/SUSY-2019-08/}{ATLAS-SUSY-2019-08}~\cite{Aad:2019vvf} & 1$\ell$ + $h(bb)$, EWK & 328 & 17\\
\href{http://cms-results.web.cern.ch/cms-results/public-results/publications/SUS-16-039/index.html}{CMS-SUS-16-039}~\cite{Sirunyan:2017lae} & multi-$\ell$, EWK  & 811 & 8\\
\href{http://cms-results.web.cern.ch/cms-results/public-results/publications/SUS-18-004/index.html}{CMS-SUS-18-004}~\cite{CMS:2021edw} & 2--3 soft $\ell$ & 33 & 4\\
\href{https://atlas.web.cern.ch/Atlas/GROUPS/PHYSICS/PAPERS/SUSY-2013-11/}{ATLAS-SUSY-2013-11}~\cite{Aad:2014vma} & 2$\ell$ ($e,\mu$), EWK & 123 &3\\
\href{https://atlas.web.cern.ch/Atlas/GROUPS/PHYSICS/PAPERS/SUSY-2013-12/}{ATLAS-SUSY-2013-12}~\cite{Aad:2014nua} & 3$\ell$ ($e,\mu,\tau$), EWK  & 2& 1\\
\href{https://atlas.web.cern.ch/Atlas/GROUPS/PHYSICS/PAPERS/SUSY-2016-07/}{ATLAS-SUSY-2016-07}~\cite{Aaboud:2017vwy} & 0$\ell$ + jets  & 598 & 0\\
\href{http://cms-results.web.cern.ch/cms-results/public-results/publications/SUS-16-048/index.html}{CMS-SUS-16-048}~\cite{CMS:2018kag} & soft OS $\ell$ & 35 & 0\\
\href{http://atlas.web.cern.ch/Atlas/GROUPS/PHYSICS/PAPERS/SUSY-2016-32/}{ATLAS-SUSY-2016-32}~\cite{Aaboud:2019trc} & HSCP  & 28 & 0\\
\href{https://atlas.web.cern.ch/Atlas/GROUPS/PHYSICS/PAPERS/SUSY-2016-24/}{ATLAS-SUSY-2016-24}~\cite{Aaboud:2018jiw} & 2--3$\ell$, EWK  & 22 & 0\\
\href{http://cms-results.web.cern.ch/cms-results/public-results/publications/SUS-20-001/index.html}{CMS-SUS-20-001}~\cite{CMS:2020bfa} & SFOS $\ell$  & 11 & 0\\
\href{https://atlas.web.cern.ch/Atlas/GROUPS/PHYSICS/PAPERS/SUSY-2013-02/}{ATLAS-SUSY-2013-02}~\cite{Aad:2014wea} & 0$\ell$ + 2--6 jets & 8 &0\\
\href{https://twiki.cern.ch/twiki/bin/view/CMSPublic/PhysicsResultsSUS13012}{CMS-SUS-13-012}~\cite{Chatrchyan:2014lfa} & jets + $\not{\!\!H}_T$ & 3 &0\\
\href{https://atlas.web.cern.ch/Atlas/GROUPS/PHYSICS/PAPERS/SUSY-2018-06/}{ATLAS-SUSY-2018-06}~\cite{Aad:2019vvi} & 3$\ell$, EWK  & 2 & 0\\
\href{http://cms-results.web.cern.ch/cms-results/public-results/publications/SUS-16-033/index.html}{CMS-SUS-16-033}~\cite{Sirunyan:2017cwe} & 0$\ell$ + jets  & 2 & 0\\
\href{http://atlas.web.cern.ch/Atlas/GROUPS/PHYSICS/PAPERS/SUSY-2015-06/}{ATLAS-SUSY-2015-06}~\cite{Aaboud:2016zdn} & 0$\ell$ + 2--6 jets & 1 & 0\\
\href{https://atlas.web.cern.ch/Atlas/GROUPS/PHYSICS/PAPERS/SUSY-2018-23/}{ATLAS-SUSY-2018-23}~\cite{ATLAS:2020qlk} & $Wh(\gamma\gamma)$, EWK & 1 & 0\\
\href{http://cms-results.web.cern.ch/cms-results/public-results/publications/SUS-16-036/index.html}{CMS-SUS-16-036}~\cite{Sirunyan:2017kqq} & 0$\ell$ + jets  & 1 & 0\\
\bottomrule
\end{tabular}
\end{table}

An overview of which analyses appear as the most sensitive ones for how many points, and how many points they exclude, is given in Table~\ref{tab:id_analysis_best}.
More concretely, the table lists all analyses from the \smodels database that appear as the `most sensitive' one for at least one point of the EWKino dataset. Also given are the number of points tested and the number of points excluded by these analyses. 
The most sensitive analyses to the gluino signal in \smodels are CMS-SUS-19-006~\cite{CMS:2019zmd} and CMS-SUS-19-010 \cite{CMS:2021beq}. They are the most sensitive analyses for 1864 points, of which 546 are excluded.\footnote{There is no equivalent ATLAS search in the \smodels database.} This can be compared to, for example, the ATLAS and CMS hadronic (0 lepton) EWKino searches, ATLAS-SUSY-2018-41~\cite{ATLAS:2021yqv}
and CMS-SUS-21-002~\cite{CMS:2022sfi}, which are the most sensitive analyses for 2247 points, of which they exclude 455. The leptonic EWKino analyses appear as most sensitive for 2593 points, excluding 254. 
It is also worth noting that, for the long-lived charginos in the EWKino dataset, the 36~\fbi disappearing track analysis from ATLAS is often more sensitive than its CMS counterpart for 101~\fbi. Concretely, ATLAS-SUSY-2016-06 is the most sensitive analysis for 871 points, excluding 274 points, while CMS-EXO-19-010 is the most sensitive analysis for 241 points, excluding 191. This is why the inclusion of this CMS analysis does not significantly improve the coverage in the compressed region.

\subsection{Effect of analysis combination} \label{sec:combination}

The multiplicity of final states and relevant analyses seen in the previous sections highlights the importance of maximizing sensitivity by leveraging all available information. In particular, the ability of \smodels to statistically combine multiple analyses 
offers a way to mitigate statistical fluctuations and compensate for limitations in individual analysis coverage~\cite{Altakach:2023tsd,MahdiAltakach:2023bdn}. This motivates a more in-depth study of the effect of such combinations, as presented in the following.

On the technical side, analysis combination requires the ability to compute likelihoods. For \smodels, this means that only analyses with efficiency map (EM) results can be used. The set of analyses we can use in this study for building global likelihoods comprises 37 analyses: 18 ATLAS and 8 CMS analyses from Run~2, as well as 6 ATLAS and 5 CMS analyses from Run~1, see the overview in appendix~\ref{app:Analyses_used}. 
However, since we do not have access to inter-analyses correlations, only analyses that are approximately uncorrelated may enter the combination. The combined likelihood is then simply the product of the likelihoods of the individual analyses. 

Generally, we assume analyses from different LHC runs, and from distinct experiments (ATLAS or CMS) to be uncorrelated. Furthermore, following \cite{Altakach:2023tsd}, we also treat analyses 
with clearly orthogonal signal regions (SRs) as approximately uncorrelated. 
This information is encoded in `combinability matrices' (one for each experiment and run) that define which analyses are considered approximately uncorrelated so they can be combined, see also appendix~\ref{app:Analyses_used}. 
We then use the {\sc Pathfinder} algorithm from Ref.~\cite{Araz:2022vtr} to determine for each point in the dataset the combination that maximizes the sensitivity to the signal; this is dubbed the `best combination'. 
Since this can be very CPU-time consuming, we only consider analyses with $\rexp>0.1$ in this process. 
Finally, if the best combination is less sensitive than the most sensitive upper limit result, it is replaced with the upper limit. From now on, we refer to this outcome as {\it the combination}.

Similar to what we found in the previous section, the results of the combination are either dominated by EWKino or by gluino constraints.
Therefore, although the gluino signal increases the overall number of excluded points, we do not expect a large increase in sensitivity from the combination of gluino and EWKino searches.
In fact, there are only 103 points for which both the most sensitive gluino analysis and the most sensitive EWKino analysis contribute significantly to the combination ($0.5<\rexp^{\rm EWKino}/\rexp^{\rm gluino}<2$). Out of these 103 points, only 25 points are excluded by the combination and were not already excluded by the most sensitive single analysis. 

\begin{figure}[!t]
    \centering
    \includegraphics[width=0.7\linewidth]{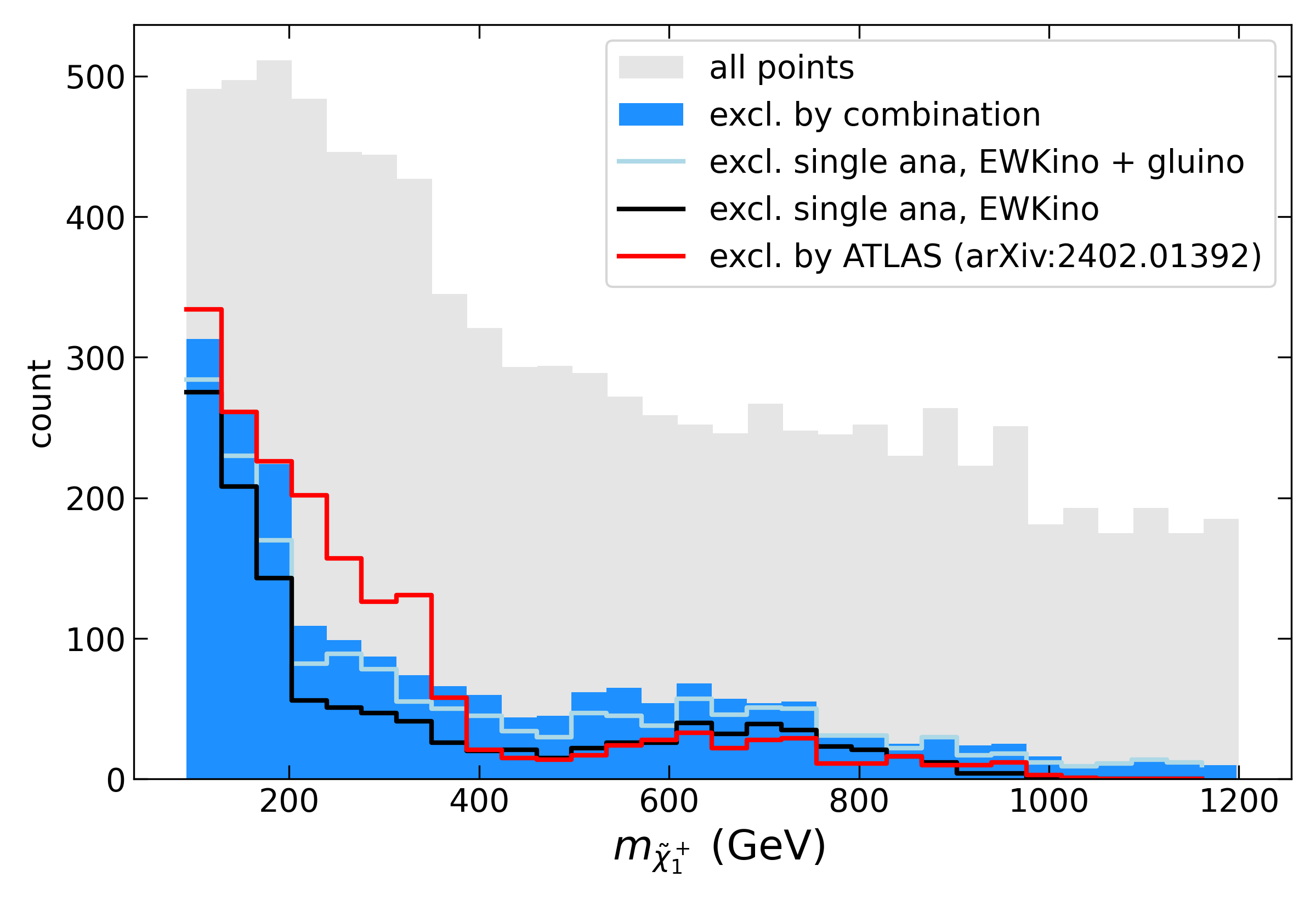}
    \caption{Number of points excluded by \smodels in different setups discussed in the text: single-analysis EWKino results (black line), EWKino and gluino results (light blue line), and statistical combination of uncorrelated analyses (filled blue histogram). Only the most sensitive analysis/combination is considered. For comparison, the red histogram shows the exclusion from \cite{ATLAS:2024qmx}. The full dataset is represented in grey.}
    \label{fig:histocomb}
\end{figure}

A larger effect is seen when the gluino is heavy and only EWKino analyses serve to constrain a point. 
In this case, we count 599 points for which at least two analyses have $0.5<\rexp<1$, and 397 (475) of them are (expected to be) excluded by the combination. 
We also notice that the combination is more relevant in the lower mass region as the number of analyses sensitive to these points increases, while only a few analyses are able to constrain heavy states. 
Overall, the observed (expected) exclusion went from 1201 (1073) points for EWKino constraints to 1696 (1479) when adding gluino constraints to 2031 (2002) points after combination. Combining analyses thus resulted in a 20\% (35\%) improvement in (expected) exclusion. 
The gain with respect to the single-analysis approach is illustrated in Fig.~\ref{fig:histocomb}.

\begin{figure}[!t]
    \hspace*{-2mm}\includegraphics[width=0.44\linewidth]{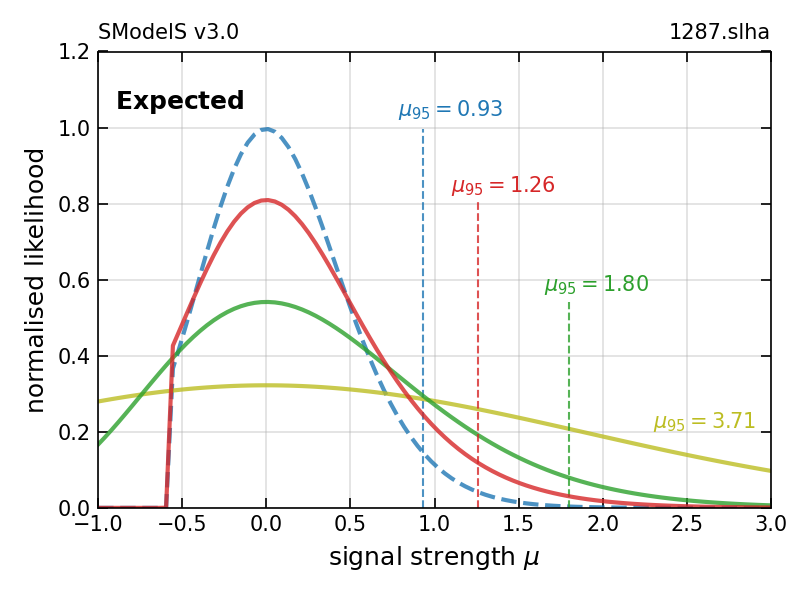}%
    \includegraphics[width=0.6\linewidth]{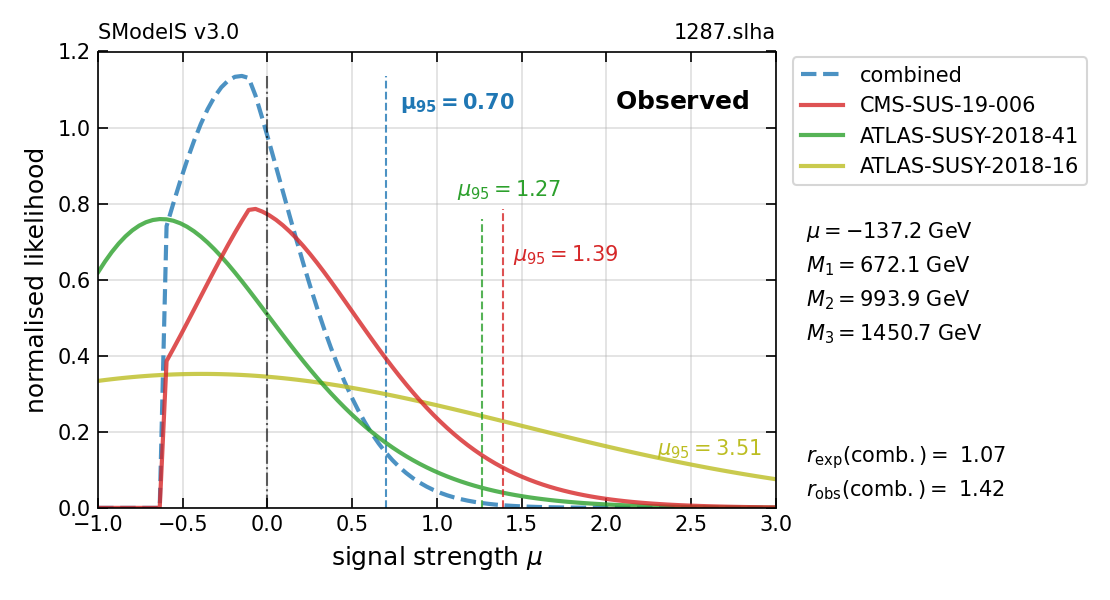}\\
    \hspace*{-2mm}\includegraphics[width=0.44\linewidth]{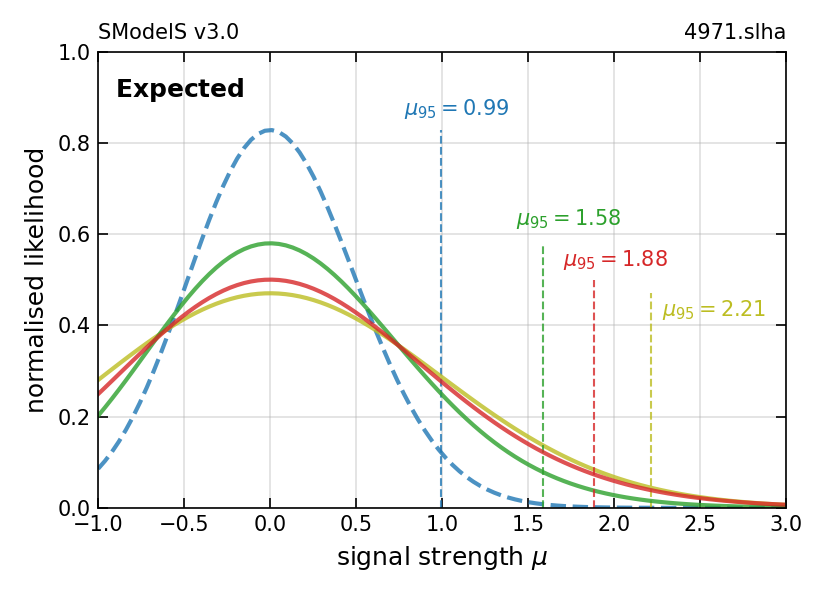}%
    \includegraphics[width=0.6\linewidth]{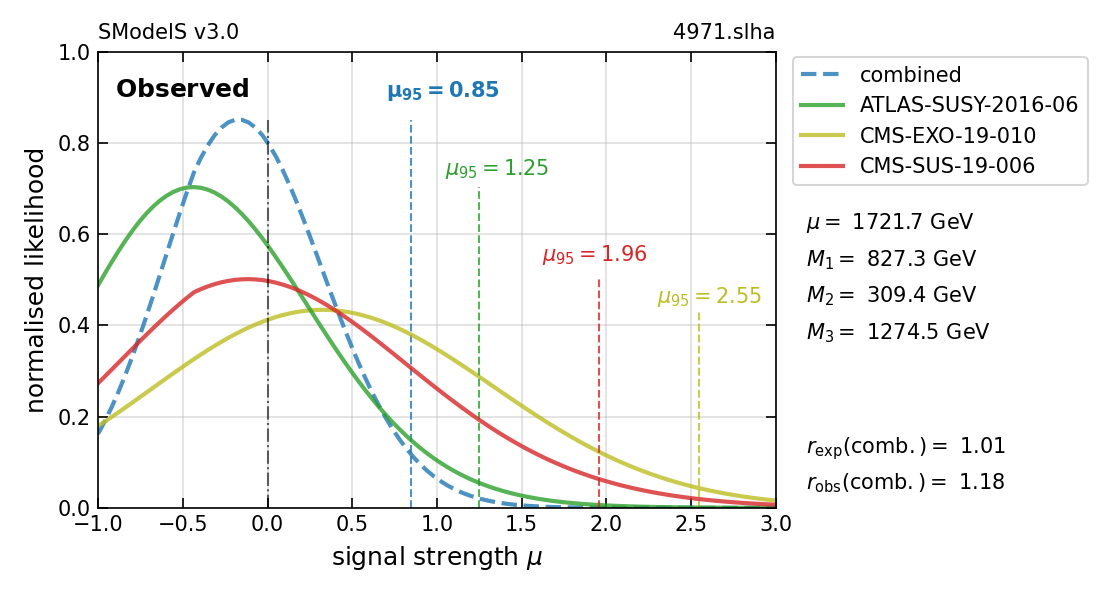}\\
    \hspace*{-2mm}\includegraphics[width=0.44\linewidth]{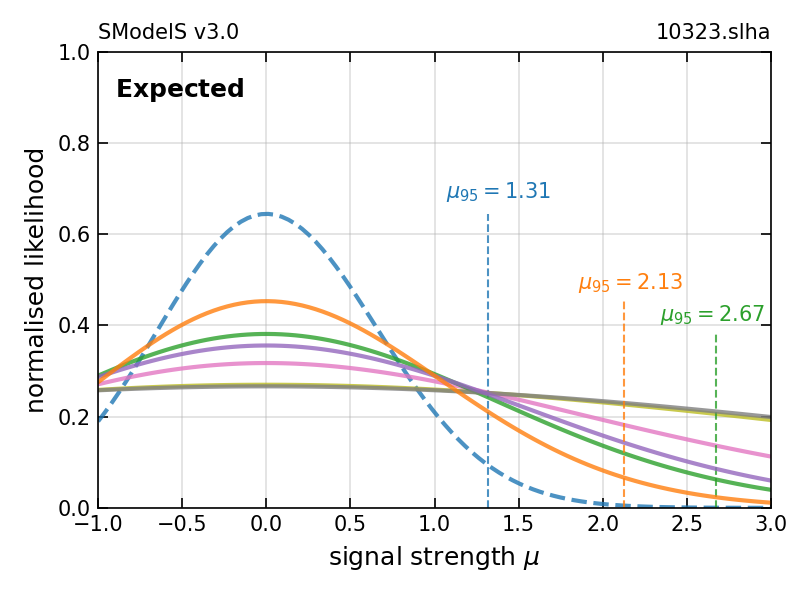}%
    \includegraphics[width=0.6\linewidth]{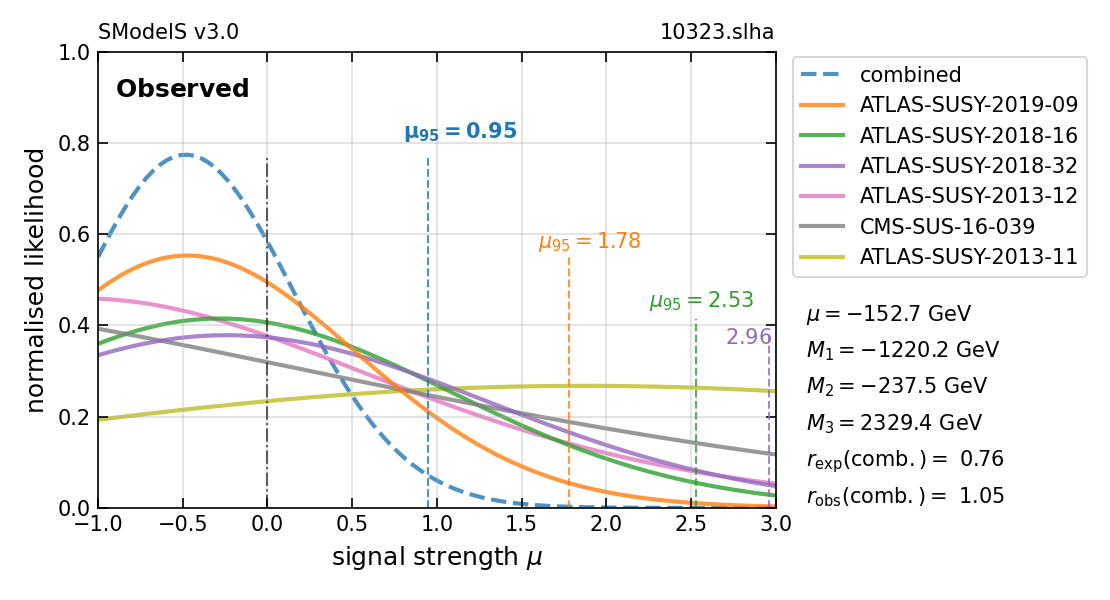}
    \caption{Normalised  expected (left) and observed (right) likelihoods as a function of the signal strength $\mu$ for three sample points: {\tt 1287} (top), {\tt 4971} (middle) and {\tt 10323} (bottom). Shown are the combined likelihoods (blue dashed lines) together with the likelihoods of each analysis entering the combination (coloured full lines). The $r$-values are given by $r=1/\mu_{95}$. The relevant SUSY Lagrangian parameters as well as the expected and observed combined $r$-values are shown on the right.}
    \label{fig:llhds}
\end{figure}

Figure~\ref{fig:llhds} shows concrete examples of likelihood combination for three sample points. 
Point {\tt 1287} (top row) features a higgsino-like LSP with a mass of 144~GeV (with $\tilde\chi^\pm_1-\tilde\chi^0_1$ and $\tilde\chi^0_2-\tilde\chi^0_1$ mass splittings below 10 GeV), a bino-like $\tilde\chi^0_3$ at about 680~GeV, and wino-like $\tilde\chi^\pm_2/\tilde\chi^0_4$ at about 1~TeV. The gluino has a mass of about 1.8~TeV and decays mainly into the higgsino-like EWKinos plus 3rd generation quarks. The relevant analyses entering the combination are two EWKino searches from ATLAS, the 0-lepton one (constraining mainly the wino production) and the soft leptons one (constraining mainly the higgsino production), and a gluino search from CMS. None of these individual analyses excludes the point, but the combination does. Also note that the combination is expected to exclude the point ($\rexp^{\rm comb}= 1/\mu_{\rm 95}^{\rm exp}= 1.07$) but the observed exclusion is even stronger ($\robs^{\rm comb}= 1/\mu_{\rm 95}^{\rm obs}=1.42$) due to a deficit of events in both ATLAS-SUSY-2018-41 and CMS-SUS-19-006.\footnote{The non-smooth behaviour of the likelihood for CMS-SUS-19-006 (red lines) at negative signal strengths comes from numerical instabilities of the covariance matrix associated to this analysis. However, since only the region with $\mu > 0$ is relevant for constraining the model points, this does not impact any of the results presented here.}

Point {\tt 4971} (middle row) is a wino-like LSP scenario with $\mnt{1}\simeq\mch{1}\simeq 340$~GeV, leading to a disappearing track signature. The gluino has a mass of about 1.6~TeV and decays to 60\% into the long-lived chargino plus $t\bar b$ or $b\bar t$. 
Thus, only a few percent of the total gluino-pair production lead to  $4t+$MET and $4b+$MET signatures, which are constrained by simplified model results in \smodels.
The bino- and higgsino-like EWKinos are too heavy to play a role. 
Neither the long-lived (disappearing track) nor the prompt search is 
sufficiently sensitive to individually exclude this point. However, when both searches are combined the point is excluded with $\robs = 1.18$. 
This illustrates the potential impact of combining prompt and long-lived results.
 
The third example, point {\tt 10323} (bottom row) features a mixed higgsino-wino scenario with a heavy gluino beyond current reach. Concretely, the higgsino-like EWKinos have masses of 143--177~GeV, the wino-like ones about 300~GeV, and the gluino about 2.7~TeV.  
The point is thus characterized by a large variety of signatures from EWKino production and decays and relatively small MET. 
Altogether six analyses, including two from Run~1, enter the combination, all having small individual $r$-values $\lesssim 0.5$. 
In principle even the combination of these six analyses is not expected to have enough sensitivity ($\rexp^{\rm comb}=0.76$), but the deficit of events with respect to the expected backgrounds pushes the combined exclusion above 1 ($\rexp^{\rm comb}=1.05$). 

\begin{figure}[!t]
    \centering
    \includegraphics[width=0.7\linewidth]{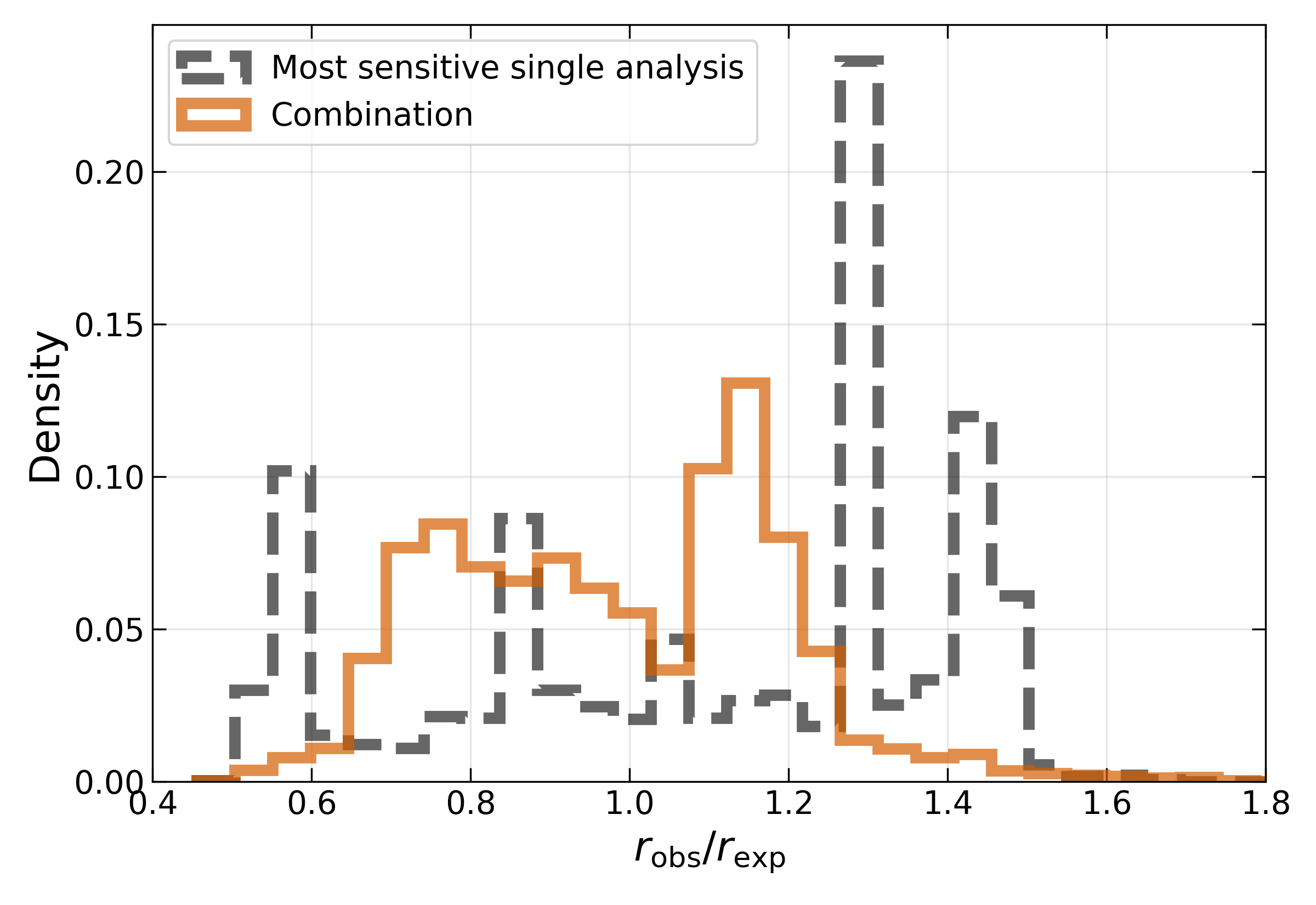}
    \caption{Ratio of $\robs/\rexp$ for the most sensitive analysis (dashed black line) and for the combination (orange). Used here is the subset of points with $\rexp>0.1$ and for which two or more analyses are combined.}
    \label{fig:histbeforeaftercomb}
\end{figure}

While the three examples above all have $\robs>\rexp$ after combination, the opposite situation also occurs. 
Note here that $\robs/\rexp<1$ means an excess, while $\robs/\rexp>1$ means an under-fluctuation was observed in the data. Generally, 
analysis combination also serves to mitigate statistical fluctuations. To illustrate this aspect, 
Fig.~\ref{fig:histbeforeaftercomb} shows the distribution of $\robs/\rexp$ before and after combination. 
In the absence of a signal, this ratio should get closer to 1 when going from the single-analysis level to the combination. This is indeed the case, as the orange histogram is more centred on 1 than the black-dashed one. Moreover, large fluctuations have decreased, as the tails of the distribution got flattened. 
Regarding the proportion of points for which \smodels output shows an excess, before combination 36\% of points have $\robs/\rexp<1$, which increases to 52\% after combination considering points with at least 2 analyses entering the combination. 

\begin{figure}[t]
    \centering
    \includegraphics[width=0.5\linewidth]{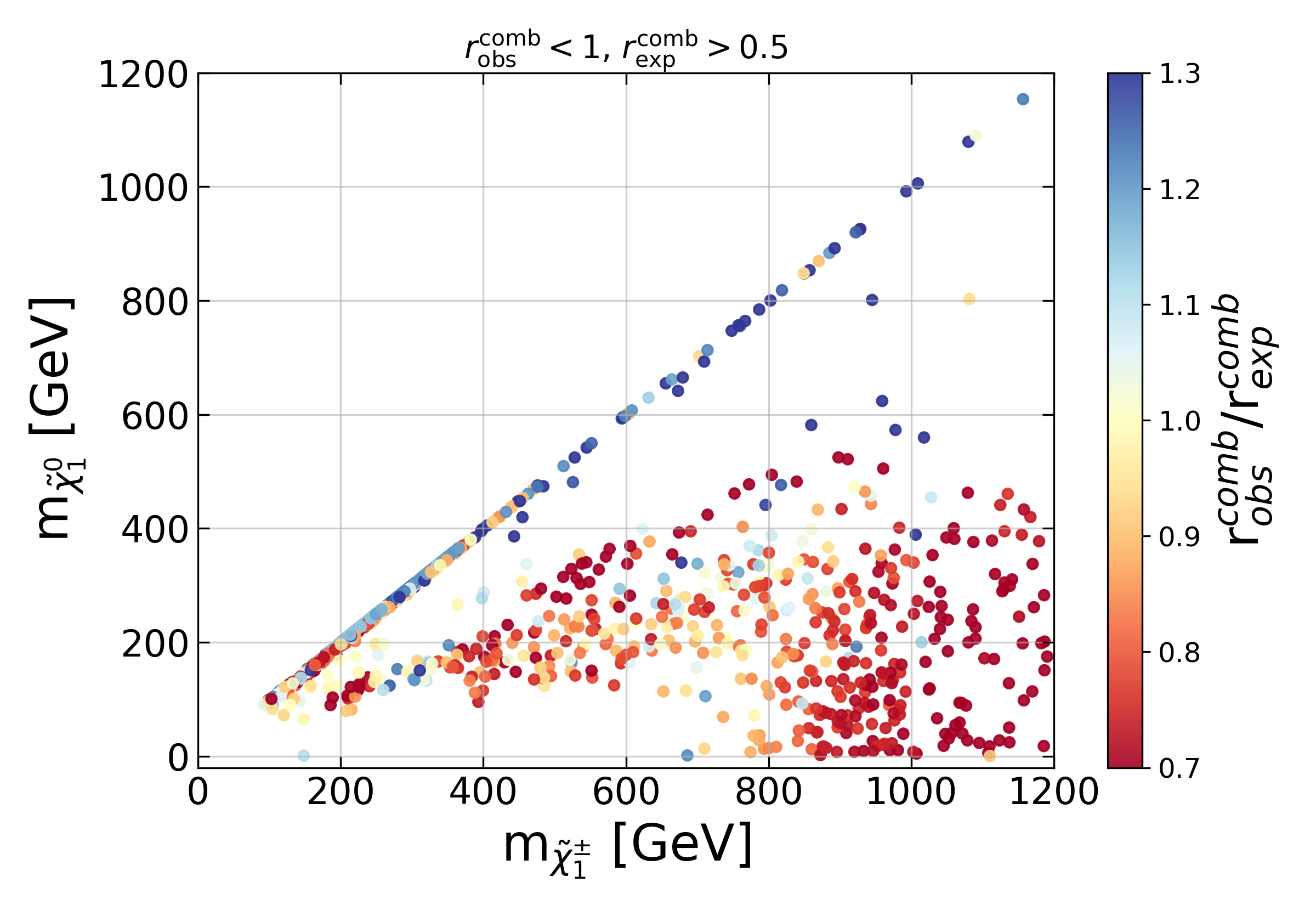}%
    \includegraphics[width=0.5\linewidth]{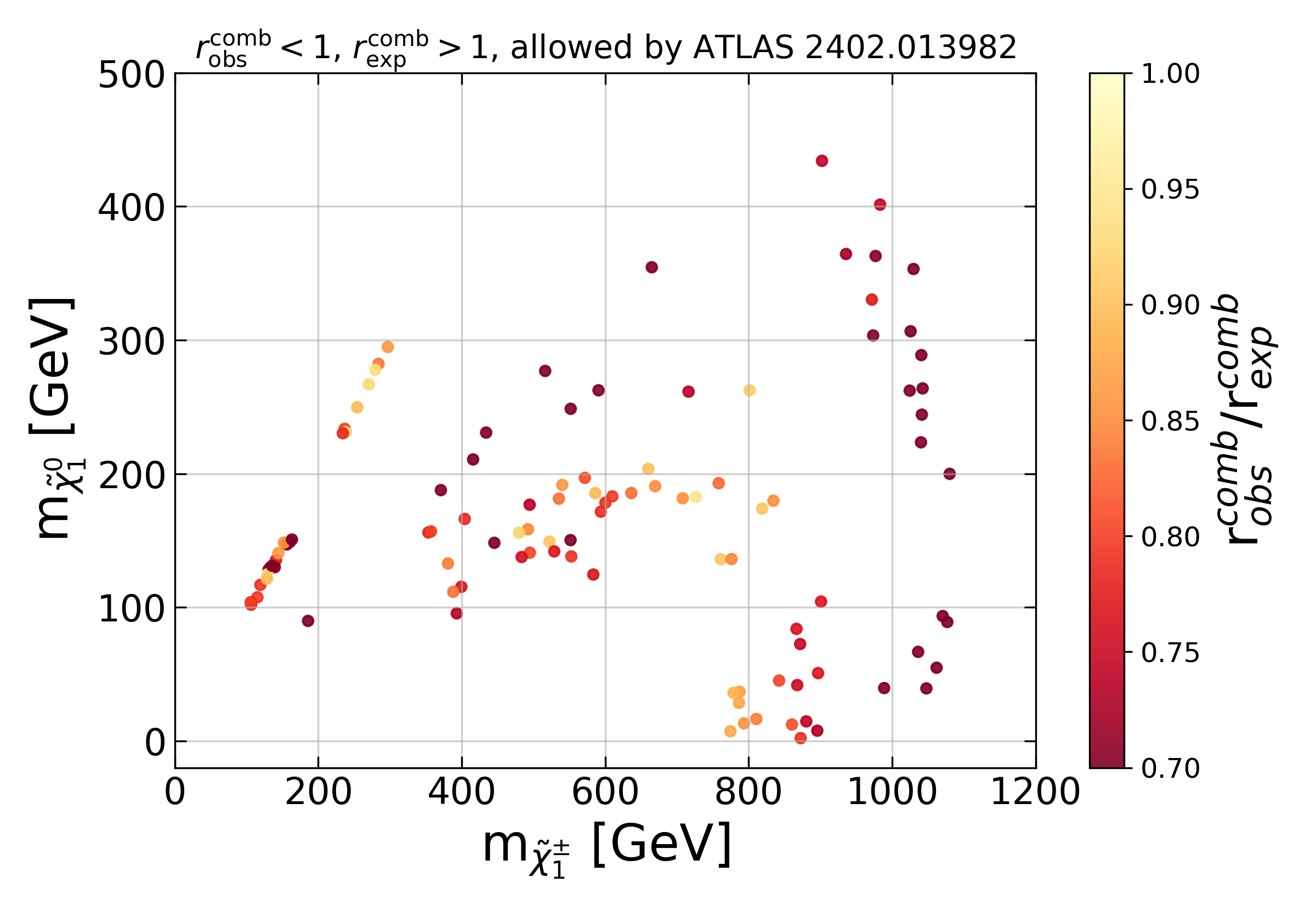}
    \caption{Scatter plots showing  $\robs^{\rm comb}/\rexp^{\rm comb}$ in the $\mnt{1}$ vs.\ $\mch{1}$ plane for points surviving exclusion ($\robs^{\rm comb}<1$). Red points represent excesses while blue points represent under-fluctuations. The left panel shows all points with $\rexp^{\rm comb}>0.5$, the right panel shows the points with $\rexp^{\rm comb}>1$ and not excluded by ATLAS~\atlasCite.}
    \label{fig:survivingexcesses}
\end{figure}

It is also interesting to see where over- and under-fluctuations are located in the parameter space. To this end, Fig.~\ref{fig:survivingexcesses} (left) shows $\robs^{\rm comb}/\rexp^{\rm comb}$ in the $\mnt{1}$ vs.\ $\mch{1}$ plane for points surviving exclusion ($\robs^{\rm comb}<1$). Only points with $\rexp^{\rm comb}>0.5$ are shown for visibility and relevance. We note that the blue points,  representing under-fluctuations in the data, are mostly concentrated along the diagonal with non-bino-like LSP and $\mch{1}>200$~GeV. Regarding excesses, represented by red points, we can identify four regions. First, there is a cluster of dark red points at high mass, $\mch{1}\gtrsim 800$~GeV. This region is constrained by the ATLAS and CMS hadronic EWKino searches~\cite{ATLAS:2021yqv,CMS:2022sfi}. As discussed also in \cite{MahdiAltakach:2023bdn,Altakach:2023tsd}, CMS records an excess (and in this region of the parameter space it is also the most sensitive analysis), while ATLAS records a small under-fluctuation, but after combination, the excess remains. 

Second, there are sizeable excesses in the almost diagonal region with $\mch{1}\approx 450$--$800$ and $\mnt{1}\approx 250$--$500$~GeV; these points mostly present wino/bino like scenarios with small mixing. 
Multiple analyses enter the combination in this region, among which the CMS hadronic search CMS-SUS-21-002~\cite{CMS:2022sfi} and the ATLAS leptonic searches ATLAS-SUSY-2019-08~\cite{Aad:2019vvf} and ATLAS-SUSY-2019-09~\cite{ATLAS:2021moa}; all  three of these report small excesses and their combination makes this more significant. 
Third, there is a small cluster of red points at $\mch{1}\approx 200$--$250$~GeV and $\mnt{1}\approx 90$--$140$~GeV. This region features light bino-higgsino scenarios with $|M_1|\lesssim|\mu|$ (with or without nearby winos: $|\mu|\lesssim|M_2|$ or $|\mu|\ll |M_2|$); the mass differences lie in the range $80~{\rm GeV}<\mnt{2}-\mnt{1}<m_h=125~{\rm GeV}$ and in most cases also $\mnt{3}-\mnt{1}<m_h$, thus leading predominately to $WZ$+MET signatures.  
Finally, the fourth region is located on the diagonal and is characterized by higgsino-like LSPs with $\mnt{1}\lesssim 200$~GeV and small $\tilde\chi^0_1$--$\tilde\chi^\pm_1$ (and $\tilde\chi^0_1$--$\tilde\chi^0_2$) mass splitting up to about 15 (20) GeV. 
This is consistent with the soft-lepton excesses reported in \cite{ATLAS:2019lng,ATLAS:2021moa,CMS:2021edw,CMS:2021cox}.\footnote{It is unfortunate that the CMS multi-lepton analysis~\cite{CMS:2021cox} does not provide any HEPData material to enable its reuse. Moreover, the CMS soft-leptons analysis~\cite{CMS:2021edw} provides only UL-type results but no efficiency maps. Therefore, neither of them can enter the combined likelihood in our study. One can expect, however, that they would enhance the excesses found for compressed spectra.} 
It will be exciting to see how the constraints in these regions will evolve with more data. 

Even more intriguing, as also discussed recently in \cite{judita}, is the question which points could have been excluded but were not because of small excesses. The set of these points is depicted in Fig.~\ref{fig:survivingexcesses} (right).  
Here, in addition to $\rexp^{\rm comb}>1$ and $\robs^{\rm comb}<1$, we require that the depicted points escape exclusion in the \atlasStudy. This is to compensate for a possible under-exclusion in \smodels due to the lack of signal topologies involving Higgs bosons in particular for ATLAS-SUSY-2018-05 (cf.\ the discussion of Fig.~\ref{fig:scatterexatlasnonexsmodels} in section~\ref{sec:comparisonAtlas}).  
While the density has thinned a lot with respect to the left panel, points remain in all four regions discussed in the previous paragraph. 
Especially tantalizing are the points with light higgsino-like LSP, which might explain the existing soft-lepton excesses.\footnote{No points with the characteristics of the bino/wino scenario advertised in \cite{Agin:2025vgn} remain from the scan.} 

\begin{figure}[!t]
    \hspace*{-2mm}\includegraphics[width=0.44\linewidth]{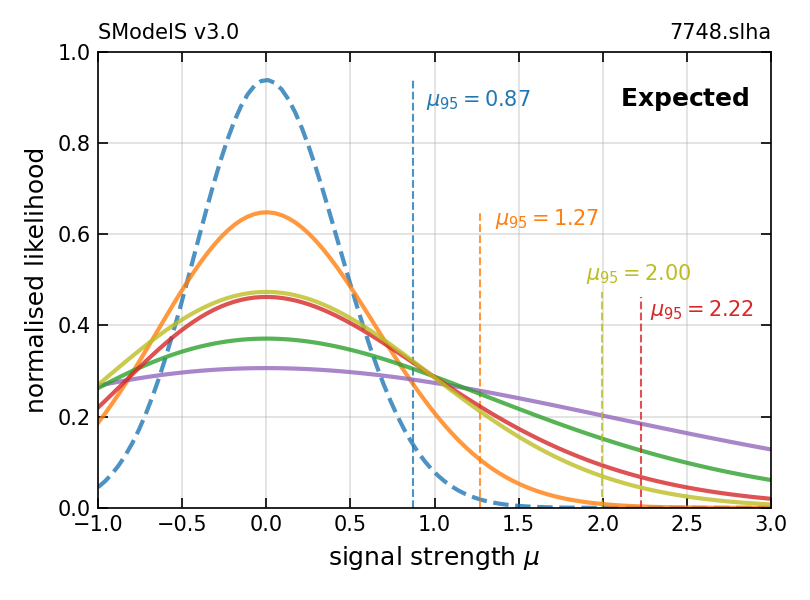}%
    \hspace*{-2mm}\includegraphics[width=0.605\linewidth]{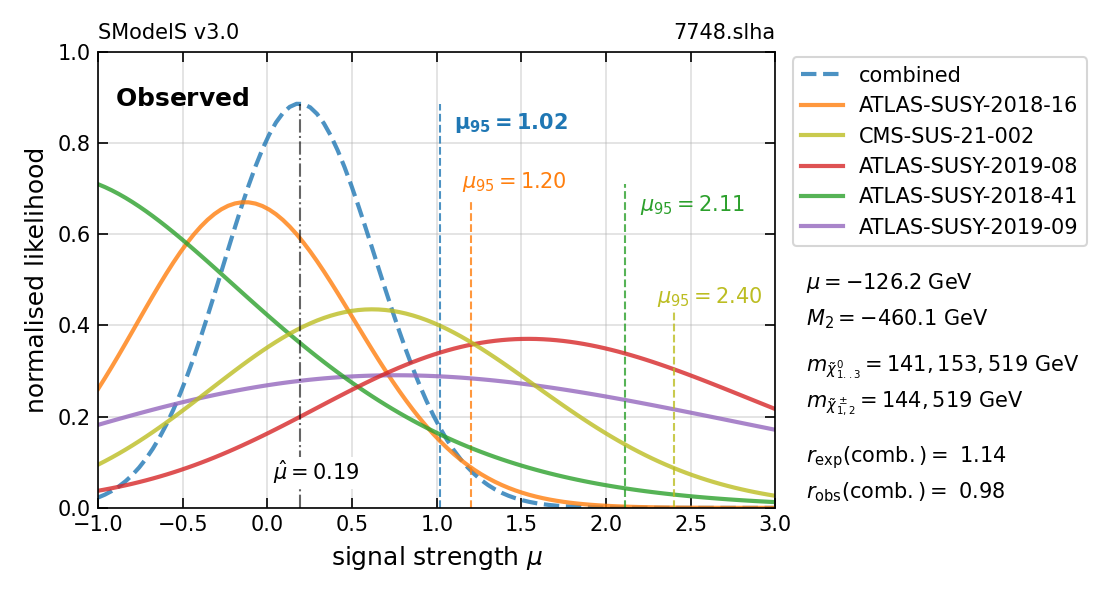}
    \caption{Normalised  expected (left) and observed (right) likelihoods as a function of the signal strength $\mu$ for point {\tt 7748}. Shown are the combined likelihoods (blue dashed lines) together with the likelihoods of each analysis entering the combination (coloured full lines). The $r$-values are given by $r=1/\mu_{95}$. The relevant SUSY Lagrangian parameters, EWKino masses, as well as the expected and observed combined $r$-values are shown in the legend on the right.}
    \label{fig:llhds7748}
\end{figure}

An explicit example for such a point is shown in Fig.~\ref{fig:llhds7748}. In this specific case, we have higgsino-like $\tilde\chi^0_1$, $\tilde\chi^\pm_1$ and $\tilde\chi^0_2$ with masses of 141, 144 and 153~GeV, respectively, and wino-like $\tilde\chi^0_3$ and $\tilde\chi^\pm_2$ with a mass of 519~GeV. Five EWKino analyses are combined in this example, four from ATLAS and one from CMS, three of which show small excesses. This leads to a combined $\robs^{\rm comb}$ just below 1, despite an expected sensitivity of $\rexp^{\rm comb}=1.14$.\footnote{For this and the other likelihood plots, we used a {\tt sigmacut} of $10^{-4}$~fb.} 

It has to be noted here that the result for the ATLAS-SUSY-2018-16~\cite{ATLAS:2019lng} analysis (orange line) in Fig.~\ref{fig:llhds7748} comes from the {\tt SR\_ewk\_2l\_high} signal region, which has a small deficit (137 expected vs.\ 134 observed) of events, while the combination of SRs in \cite{ATLAS:2019lng} shows an overall excess. Unfortunately, however, there is no statistical model available that would allow the results from the four higgsino-specific EMs to be combined in \smodels. Yet, in \cite{ATLAS:scandata}, ATLAS reports an observed CL$_{s}$ value of 0.2172 for this point, versus an expected one of 0.00857. We can therefore conclude that the tension between $\robs^{\rm comb}$ and $\rexp^{\rm comb}$ for point {\tt 7748} should be even larger than shown in Fig.~\ref{fig:llhds7748}. 
Furthermore, as said in footnote~10, two more CMS analyses should be relevant and also have potentially consistent excesses, but they cannot be used here because they not provide the needed information.   
All in all, as also highlighted in \cite{Agin:2023yoq,Chakraborti:2024pdn,Agin:2024yfs,Martin:2024pxx,Agin:2025vgn}, 
the situation is tantalizing and it will be highly exciting to see whether the data from Run~3 further strengthen the existing small excesses or exclude them.

\section{Summary and Conclusions}\label{sec:conclusions}

We presented a detailed comparison of the constraints on the EWKino sector of the pMSSM derived with SModelS~v3.0 against the results presented by the \atlasStudy. 
Following the same “single-analysis” approach as the ATLAS study (i.e.\ taking
the constraint of the most sensitive analysis for each parameter point), we first discussed which regions and scenarios are well covered by \smodels as compared to ATLAS, and where additional and/or updated results would be needed. Along the way, we could also clarify some inconsistencies in the ATLAS dataset regarding too long-lived charginos. 
We then extended our analysis to the full SModelS database, that is including also CMS results and in particular constraints from gluino production, and discussed how this increases the overall coverage of the parameter space. 

Finally, we showed what can be gained by the statistical combination of (approximately) uncorrelated analyses. 
Here, we considered a total of 37 analyses from the \smodels database, for which likelihoods can be constructed thanks to simplified-model efficiency maps. 
We used the {\sc Pathfinder} algorithm to determine for each point in the dataset the combination of analyses that maximizes the sensitivity to the signal. Our results demonstrated not only the gain in constraining power through a global likelihood analysis, but also the mitigation of statistical fluctuations. 
A comparison of the number of excluded points for bino-like LSP and non-bino-like LSP for each step in our analysis is given in 
Table~\ref{tab:summary}. 
Last but not least, we also highlighted the landscape of small excesses, consistent with all constraints in the database, that emerges from a global likelihood analysis.

\begin{table}[t!]\centering
\begin{tabular}{l|c|c}
 & bino-like LSP & non-bino LSP \\
 \hline
Total number of points after filtering & 3034 & 5919 \\
\hline
Number of points excluded by: & & \\
\quad 1. ATLAS in~\cite{ATLAS:2024qmx}  & 529 (17\%) & 1271 (21\%)\\
\quad 2. ATLAS EWKino results in \smodels &575 (19\%) & 687 (12\%) \\
\quad 3. all EWKino results in \smodels & 539 (18\%) & 662 (11\%)\\ 
\quad 4. full \smodels database & 666 (22\%) & 1030 (17\%)\\
\quad 5. full DB with analysis combination & 847 (28\%) & 1184 (20\%)\\
\bottomrule
\end{tabular}
\caption{Comparison of the number of excluded points for bino-like LSP and non-bino-like LSP. Note that for single-analysis results (cases 1--4), the most sensitive, not the most constraining, analysis is considered.}\label{tab:summary}
\end{table}

We conclude that the combination of analyses in \smodels significantly enhances the robustness and sensitivity of statistical  tests in the pMSSM scan. It mitigates statistical fluctuations, especially in regions with dense analysis coverage, and enables more realistic interpretations when combining ATLAS and CMS results. However, the improvement is still limited by the availability of efficiency maps and the granularity of simplified model interpretations. Extending the \smodels database to include more recent long-lived particle searches and Higgs-mediated decay topologies would further strengthen its sensitivity, especially for compressed or higgsino-like spectra where the current sensitivity remains limited. To make this and further extensions possible, we urge the ATLAS and CMS collaborations to more systematically provide efficiency maps for the relevant simplified models.

Last but not least, the combination of analyses also allows one to elucidate particularly interesting regions of small but consistent excesses, that should be specially probed with more data. 
We are thus eagerly awaiting the corresponding results from Run~3 of the LHC, and we hope that they will be published in a way that enables their wide reuse. \\

We end this paper by noting that we have not yet included the mono-jet results from~\cite{ATLAS:2021kxv,CMS:2021far}, which also show small excesses. Their implementation and usage in \smodels for EWKino production is available from v3.1 onwards. Studying their impact in a global likelihood analysis is left for future work.

\paragraph{Data availability} 
\smodels is public software distributed under the GPLv3 licence. 
All data and results presented in this work, together with jupyter notebooks to reproduce the paper plots, are publicly available on {\sc Zenodo}~\cite{zenodoEntry}.

\section*{Acknowledgements}
We thank Werner Porod and Jonas W\"urzinger for discussions, and the ATLAS collaboration for making their complete pMSSM scan and detailed information on their results publicly available. LC particularly thanks Sahana Narasimha and Timoth\'ee Pascal for discussions on analysis combination and \smodels in general. 

\paragraph{Funding information}
The work presented in this paper was supported in part by the French Agence Nationale de la Recherche
under grants ANR-21-CE31-0023 (SLDNP) and ANR-23-CHRO-0006 (OpenMAPP),  
by FAPESP under grants no.~2018/25225-9 and 2021/01089-1, and by the Austrian Science Fund FWF under grant number I5767-N.

\begin{appendix}
\section{Discussion of the ATLAS-SUSY-2018-41 implementation}
\label{app:ATLAS search}

In section~\ref{sec:comparisonAtlas} we noted that 166 of the 222 points over-excluded by \smodels compared to ATLAS 
(i.e.\ excluded by the most sensitive ATLAS analysis in \smodels but not excluded by ATLAS in~\cite{ATLAS:2024qmx}) are excluded by the ATLAS-SUSY-2018-41 analysis in \smodels. It is thus relevant to take a closer look at this implementation.

The ATLAS search for charginos and neutralinos in fully hadronic final states, ATLAS-SUSY-2018-41~\cite{ATLAS:2021yqv}, was implemented in \smodels in v2.2 and updated in v2.3~\cite{MahdiAltakach:2023bdn}. 
The analysis uses large-radius jets and jet substructure information to identify boosted $W$, $Z$ or Higgs bosons.
Two orthogonal SR categories, 4Q and 2B2Q, are defined for final states with four light-flavor jets ($qqqq$) and final states with two $b$-jets and two light-flavor jets ($bbqq$), respectively; events with leptons are vetoed. Multiple SRs are defined in each category to target final states from different combinations of Standard Model bosons. 
Moreover, three SRs are inclusive in $V=W,Z$: SR-2B2Q-VZ, SR-2B2Q-Vh and SR-4Q-VV.

The efficiency maps implemented in \smodels concern these latter inclusive SRs. Concretely, they are based on the acceptance and efficiency tables provided on \hepdata~\cite{hepdata.104458.v1} for the SR-2B2Q-VZ, SR-2B2Q-Vh and SR-4Q-VV SRs for different signal hypotheses: $WW$, $WZ$, $Wh$, $ZZ$, $Zh$, $hh$~(+~MET). 
Since the inclusive SRs are described as statistically independent in \cite{ATLAS:2021yqv}, they are combined in \smodels by means of a diagonal covariance matrix. 
This combination of signal contributions in different SRs is important to achieve a good coverage, but leads to a slight over-exclusion for some simplified model scenarios, in particular for $\tilde\chi^\pm_1\tilde\chi^0_2\to WZ+{\rm MET}$. This is illustrated in, e.g.,~Fig.~6 in~\cite{Altakach:2023tsd}. 

In an update of the \hepdata entry, ATLAS later provided the full {\sc HistFactory} statistical model for the analysis~\cite{hepdata.104458.v3.r9}. This should allow one to exactly reproduce the statistical evaluation in~\cite{ATLAS:2021yqv}. Moreover, signal efficiencies for all SRs should in principle be extractable from the signal patchsets included in~\cite{hepdata.104458.v3.r9}. Unfortunately, in terms of ``pure'' simplified models, as needed by \smodels, these patchsets cover only  the $WW$, $WZ$, $Wh$ +~MET signals, but not the $ZZ$, $Zh$, $hh$ ones. 
Without the full set of final states, however, the coverage of generic signals, like prevalent in the pMSSM, becomes too poor. Therefore, \smodels stayed with the original implementation of the inclusive SRs and the diagonal covariance matrix.\footnote{The {\sc HistFactory} model is not useable for the inclusive SRs.} 

In the subset of the 166 points over-excluded by the ATLAS-SUSY-2018-41 analysis in \smodels in this study, the main contribution typically comes from the $WZ$+MET topology. The $WZ$+MET signal is picked up in two of the inclusive SRs: SR-2B2Q-VZ and SR-4Q-VV. 
One can thus ask whether the apparent over-exclusion could be mitigated by using the $WZ$+MET signal from only one of these SRs instead of both. 
The effect of doing so is shown in Table~\ref{table:2018-41} and compared to the result from the full implementation. 
Used here is the subset of points for which ATLAS-SUSY-2018-41 is the most sensitive analysis in \smodels or in the ATLAS study~\cite{ATLAS:2024qmx}. 

It turns out that using the $WZ$+MET signal only from SR-4Q-VV mitigates the over-exclusion for 22 points, but leads to an additional 7 points that are excluded by ATLAS but not by \smodels. The total improvement thus amounts to only 15 points. 
Using the $WZ$+MET signal only from SR-2B2Q-VZ, reduces the number of over-excluded points by 90, but increases the number of under-excluded points by 28. Bearing in mind that SR-4Q-VV is the more sensitive region over most of the chargino/neutralino mass plane, especially in the high-mass region, while SR-2B2Q-VZ comes into play mostly for low masses, neither solution seems clearly preferable over the existing official implementation. We thus keep the ATLAS-SUSY-2018-41 implementation in \smodels as is. 

\begin{table}[!t]\centering
\begin{tabular}{l|c|c}
 & excl. SModelS not ATLAS & excl. ATLAS not SModelS \\
\hline
official database  & 166 & 113\\
SR-4Q-VV  & 144 & 120 \\
SR2B2Q-VZ & 76 & 141\\ 
\bottomrule
\end{tabular}
\caption{Comparison of the number of points excluded in \smodels but not in \cite{ATLAS:2024qmx} with the number of points excluded by \cite{ATLAS:2024qmx} but not in \smodels, for the official implementation of ATLAS-SUSY-2018-41 in \smodels, and two alternative versions with $WZ$+MET efficiency maps kept in only one of two SRs.}
\label{table:2018-41}
\end{table}

\section{Analyses from the \smodels database used in this work}\label{app:Analyses_used}

\begin{figure}[!h]
    \includegraphics[width=0.48\linewidth]{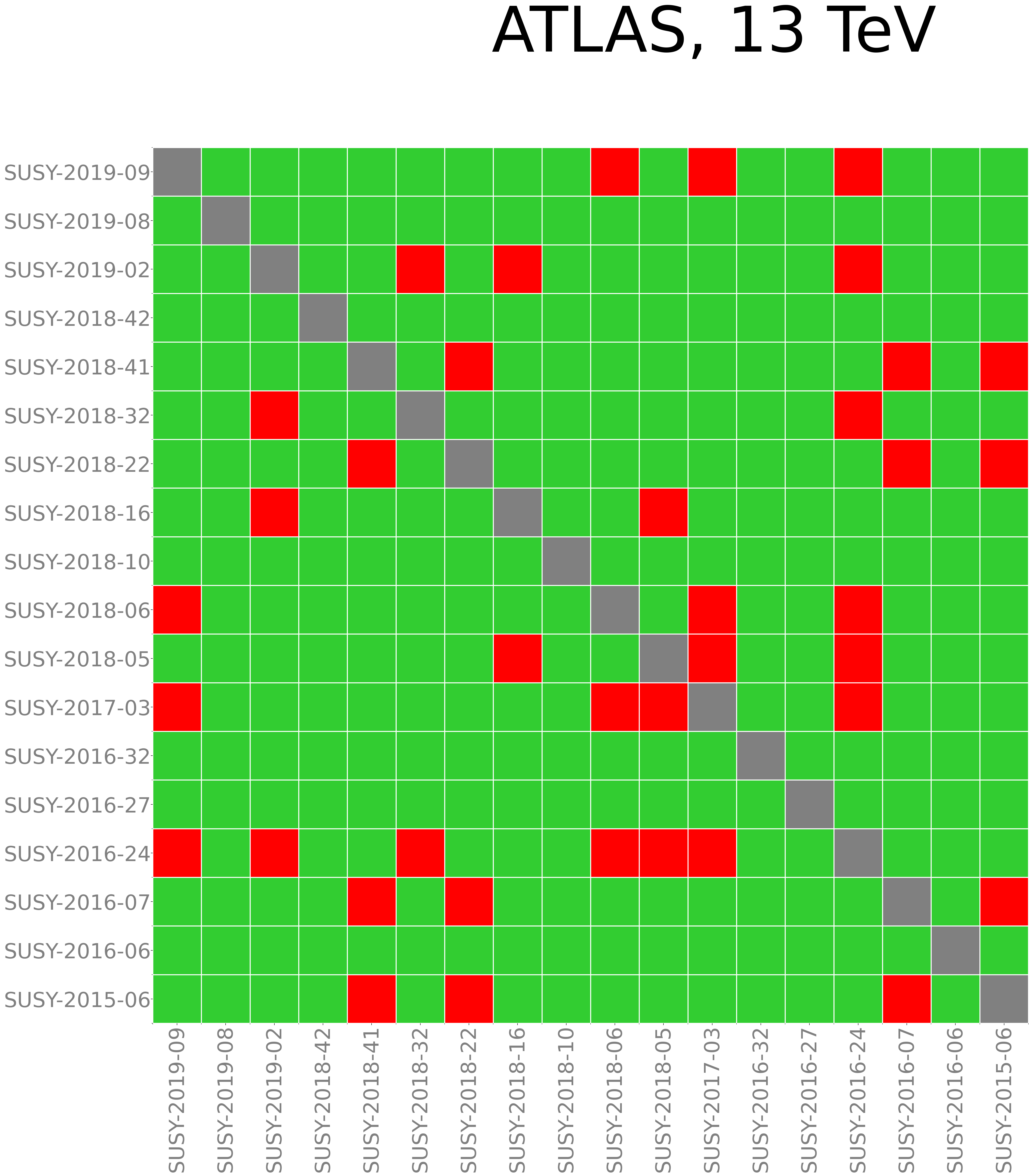} 
    \includegraphics[width=0.48\linewidth]{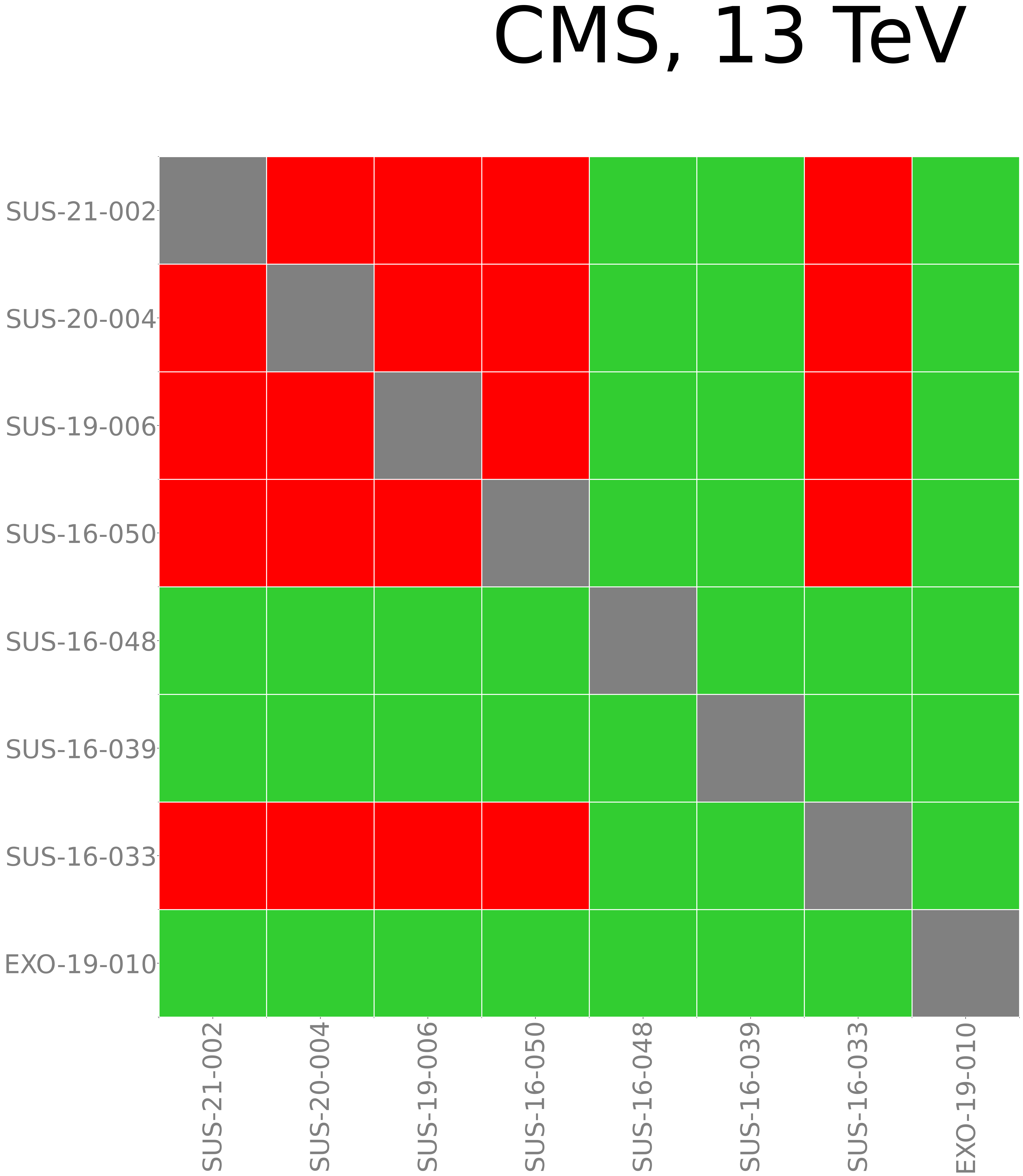}
    \caption{Combination matrices for ATLAS and CMS 13~TeV analyses; green means combinable, red means not combinable.}
    \label{fig:matrix_comb13}
\end{figure}

\begin{figure}[!h]
    \includegraphics[width=0.48\linewidth]{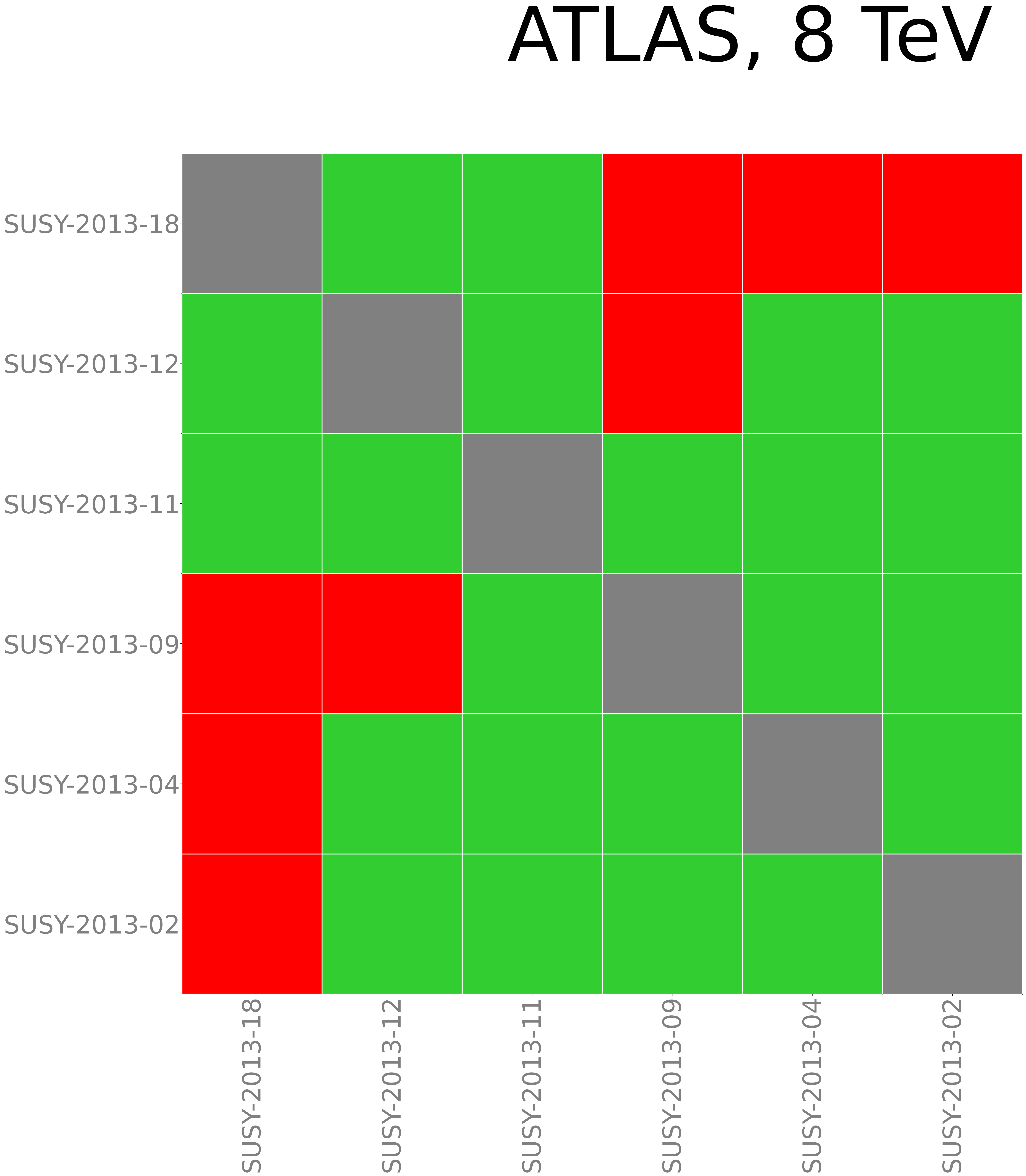}
    \includegraphics[width=0.48\linewidth]{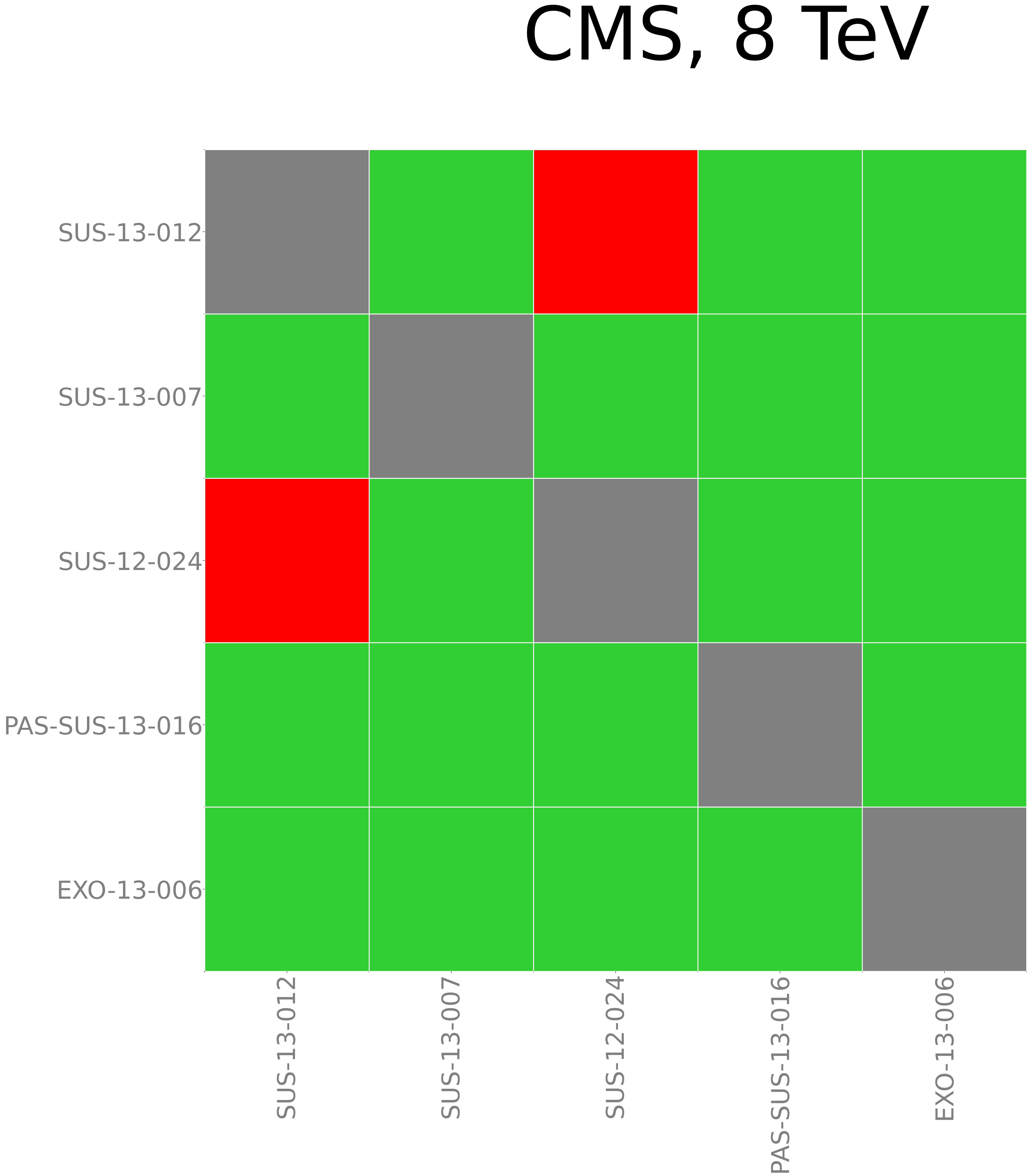}%
    \caption{Combination matrices for ATLAS and CMS 8~TeV analyses; green means combinable, red means not combinable.}
    \label{fig:matrix_comb8}
\end{figure}

We here give an overview of all analyses from the \smodels database~3.0.0 that were used in this work. Tables~\ref{tab:analist_13tev} and \ref{tab:analist_8tev} list all the analyses that constrained at least one point of the EWKino dataset. Note that this includes all results for EWKino or gluino production, as long as an expected $r$-value can be computed (results with efficiency maps or, in case of cross section upper limit maps, both expected and observed upper limits). In turn, this excludes all analyses for which only observed cross section upper limits are available. 
The information which analyses are considered approximately uncorrelated so they can be combined is given in Figs.~\ref{fig:matrix_comb13} and~\ref{fig:matrix_comb8}.

A list of {\it all} analyses in the latest \smodels database is available online at \url{https://smodels.github.io/docs/ListOfAnalyses}.

\begin{table}[h!]\centering
\caption{ATLAS and CMS Run~2 ($\sqrt{s}=13$~TeV) analyses from the \smodels v3.0.0 database used in this study. A checkmark in the column {\bf UL} means that observed and expected ULs are available; a checkmark in the column {\bf EM} means that efficiency maps are available. The column {\bf comb.}\ specifies whether and how signal regions (SRs) are combined for EM results: ``pyhf'' means a \hf model is used through interface with \pyhf~\cite{Heinrich:2021gyp,pyhf}; SLv1 (SLv2) means that a covariance matrix is used in the Simplified Likelihood scheme of \cite{CMS:SL} (\!\!\cite{Buckley:2018vdr}); a -- means that only the most sensitive SR is used.} \label{tab:analist_13tev} 
\begin{tabular}{|l|l|c|c|c|c|c|c|c|}
\hline 
{\bf Analysis ID} & {\bf Short Description} & {\bf $\mathcal{L}$ [fb$^{-1}$] } & {\bf UL} & {\bf EM}& {\bf comb.}\\
\hline
\hline
\href{http://atlas.web.cern.ch/Atlas/GROUPS/PHYSICS/PAPERS/SUSY-2015-06/}{ATLAS-SUSY-2015-06}~\cite{Aaboud:2016zdn} & 0$\ell$ + 2--6 jets & 3.2 &   & \checkmark&  -- \\
\href{https://atlas.web.cern.ch/Atlas/GROUPS/PHYSICS/PAPERS/SUSY-2016-06/}{ATLAS-SUSY-2016-06}~\cite{Aaboud:2017mpt} & disappearing track & 36.1 &   & \checkmark&  -- \\
\href{https://atlas.web.cern.ch/Atlas/GROUPS/PHYSICS/PAPERS/SUSY-2016-07/}{ATLAS-SUSY-2016-07}~\cite{Aaboud:2017vwy} & 0$\ell$ + jets & 36.1 &   & \checkmark& --  \\
\href{https://atlas.web.cern.ch/Atlas/GROUPS/PHYSICS/PAPERS/SUSY-2016-24/}{ATLAS-SUSY-2016-24}~\cite{Aaboud:2018jiw} & 2--3$\ell$, EWK & 36.1 &   & \checkmark&  -- \\
\href{https://atlas.web.cern.ch/Atlas/GROUPS/PHYSICS/PAPERS/SUSY-2016-27/}{ATLAS-SUSY-2016-27}~\cite{Aaboud:2018doq} & jets + $\gamma$ & 36.1 &   & \checkmark& --  \\
\href{http://atlas.web.cern.ch/Atlas/GROUPS/PHYSICS/PAPERS/SUSY-2016-32/}{ATLAS-SUSY-2016-32}~\cite{Aaboud:2019trc} & HSCP & 31.6 & \checkmark & \checkmark& --  \\
\href{https://atlas.web.cern.ch/Atlas/GROUPS/PHYSICS/PAPERS/SUSY-2017-03/}{ATLAS-SUSY-2017-03}~\cite{Aaboud:2018sua} & multi-$\ell$ EWK & 36.1 &   & \checkmark&  -- \\
\href{https://atlas.web.cern.ch/Atlas/GROUPS/PHYSICS/PAPERS/SUSY-2018-05/}{ATLAS-SUSY-2018-05}~\cite{ATLAS:2022zwa} & 2$\ell$ + jets, EWK & 139.0 & \checkmark & \checkmark& pyhf \\
\href{https://atlas.web.cern.ch/Atlas/GROUPS/PHYSICS/PAPERS/SUSY-2018-06/}{ATLAS-SUSY-2018-06}~\cite{Aad:2019vvi} & 3$\ell$, EWK & 139.0 & \checkmark & \checkmark&  -- \\
\href{https://atlas.web.cern.ch/Atlas/GROUPS/PHYSICS/PAPERS/SUSY-2018-10/}{ATLAS-SUSY-2018-10}~\cite{ATLAS:2021twp} & 1$\ell$ + jets & 139.0 &   & \checkmark& --  \\
\href{https://atlas.web.cern.ch/Atlas/GROUPS/PHYSICS/PAPERS/SUSY-2018-16/}{ATLAS-SUSY-2018-16}~\cite{ATLAS:2019lng} & 2 soft $\ell$ + jets, EWK & 139.0 & \checkmark & \checkmark& pyhf \\
\href{https://atlas.web.cern.ch/Atlas/GROUPS/PHYSICS/PAPERS/SUSY-2018-22/}{ATLAS-SUSY-2018-22}~\cite{ATLAS:2020syg} & multi-jets & 139.0 & \checkmark & \checkmark& pyhf \\
\href{https://atlas.web.cern.ch/Atlas/GROUPS/PHYSICS/PAPERS/SUSY-2018-23/}{ATLAS-SUSY-2018-23}~\cite{ATLAS:2020qlk} & $Wh(\gamma\gamma)$, EWK & 139.0 & \checkmark &  &   \\
\href{https://atlas.web.cern.ch/Atlas/GROUPS/PHYSICS/PAPERS/SUSY-2018-32/}{ATLAS-SUSY-2018-32}~\cite{Aad:2019vnb} & 2 OS $\ell$ & 139.0 &   & \checkmark& pyhf \\
\href{https://atlas.web.cern.ch/Atlas/GROUPS/PHYSICS/PAPERS/SUSY-2018-41/}{ATLAS-SUSY-2018-41}~\cite{ATLAS:2021yqv} & hadr. EWK & 139.0 & \checkmark & \checkmark& SLv1 \\
\href{https://atlas.web.cern.ch/Atlas/GROUPS/PHYSICS/PAPERS/SUSY-2018-42/}{ATLAS-SUSY-2018-42}~\cite{ATLAS:2022pib} & charged LLPs, dE/dx & 139.0 & \checkmark & \checkmark& --  \\
\href{https://atlas.web.cern.ch/Atlas/GROUPS/PHYSICS/PAPERS/SUSY-2019-02/}{ATLAS-SUSY-2019-02}~\cite{ATLAS:2022hbt} & 2 soft $\ell$, EWK & 139.0 &   & \checkmark& SLv1 \\
\href{https://atlas.web.cern.ch/Atlas/GROUPS/PHYSICS/PAPERS/SUSY-2019-08/}{ATLAS-SUSY-2019-08}~\cite{Aad:2019vvf} & 1$\ell$ + $h(bb)$, EWK & 139.0 &   & \checkmark& pyhf \\
\href{https://atlas.web.cern.ch/Atlas/GROUPS/PHYSICS/PAPERS/SUSY-2019-09/}{ATLAS-SUSY-2019-09}~\cite{ATLAS:2021moa} & 3$\ell$, EWK & 139.0 & \checkmark & \checkmark& pyhf \\
\hline
\hline
\href{http://cms-results.web.cern.ch/cms-results/public-results/publications/EXO-19-010/}{CMS-EXO-19-010}~\cite{CMS:2020atg} & disappearing track & 101.0 &   & \checkmark&  -- \\
\href{https://cms-results.web.cern.ch/cms-results/public-results/publications/SUS-16-009/}{CMS-SUS-16-009}~\cite{Khachatryan:2017rhw} & 0$\ell$ + jets, top tag & 2.3 & \checkmark &  &   \\
\href{http://cms-results.web.cern.ch/cms-results/public-results/publications/SUS-16-033/index.html}{CMS-SUS-16-033}~\cite{Sirunyan:2017cwe} & 0$\ell$ + jets & 35.9 & \checkmark & \checkmark&  -- \\
\href{http://cms-results.web.cern.ch/cms-results/public-results/publications/SUS-16-036/index.html}{CMS-SUS-16-036}~\cite{Sirunyan:2017kqq} & 0$\ell$ + jets & 35.9 & \checkmark &  &   \\
\href{http://cms-results.web.cern.ch/cms-results/public-results/publications/SUS-16-039/index.html}{CMS-SUS-16-039}~\cite{Sirunyan:2017lae} & multi-$\ell$, EWK & 35.9 &  & \checkmark& SLv1 \\
\href{http://cms-results.web.cern.ch/cms-results/public-results/publications/SUS-16-048/index.html}{CMS-SUS-16-048}~\cite{CMS:2018kag} & soft OS $\ell$ & 35.9 &   & \checkmark& SLv1 \\
\href{http://cms-results.web.cern.ch/cms-results/public-results/publications/SUS-16-050/index.html}{CMS-SUS-16-050}~\cite{Sirunyan:2017pjw} & 0$\ell$ + top tag & 35.9 & \checkmark & \checkmark& SLv1 \\
\href{https://cms-results.web.cern.ch/cms-results/public-results/publications/SUS-18-002/}{CMS-SUS-18-002}~\cite{Sirunyan:2019hzr} & $\gamma$ + ($b$-)jets, top tag & 35.9 & \checkmark &  &   \\
\href{http://cms-results.web.cern.ch/cms-results/public-results/publications/SUS-18-004/index.html}{CMS-SUS-18-004}~\cite{CMS:2021edw} & 2--3 soft $\ell$ & 137.0 & \checkmark &  &   \\
\href{http://cms-results.web.cern.ch/cms-results/public-results/publications/SUS-19-006/index.html}{CMS-SUS-19-006}~\cite{Sirunyan:2019ctn} & 0$\ell$ + jets, $\not{\!\!H}_T$ & 137.0 & \checkmark & \checkmark& SLv1 \\
\href{http://cms-results.web.cern.ch/cms-results/public-results/publications/SUS-19-008/index.html}{CMS-SUS-19-008}~\cite{CMS:2020cpy} & 2--3$\ell$ + jets & 137.0 & \checkmark &  &   \\
\href{http://cms-results.web.cern.ch/cms-results/public-results/publications/SUS-19-010/index.html}{CMS-SUS-19-010}~\cite{CMS:2021beq} & jets + top- and $W$-tag & 137.0 & \checkmark &  &   \\
\href{http://cms-results.web.cern.ch/cms-results/public-results/publications/SUS-20-001/index.html}{CMS-SUS-20-001}~\cite{CMS:2020bfa} & SFOS $\ell$ & 137.0 & \checkmark &  &   \\
\href{http://cms-results.web.cern.ch/cms-results/public-results/publications/SUS-20-004/}{CMS-SUS-20-004}~\cite{CMS:2022vpy} & 2 $h(bb)$, EWK & 137.0 & \checkmark & \checkmark& SLv2 \\
\href{http://cms-results.web.cern.ch/cms-results/public-results/publications/SUS-21-002/}{CMS-SUS-21-002}~\cite{CMS:2022sfi} & hadr. EWK & 137.0 & \checkmark & \checkmark& SLv1 \\
\hline
\end{tabular}\end{table}

\begin{table}[h!]\centering
\caption{ATLAS and CMS Run~1 ($8$~TeV) analyses from the \smodels v3.0.0 database used in this study. The column {\bf comb.} from Table~\ref{tab:analist_13tev} is omitted because none of the Run~1 analyses provides any information on background correlations. 
}\label{tab:analist_8tev}
\begin{tabular}{|l|l|c|c|c|c|c|c|}
\hline 
{\bf Analysis ID} & {\bf Short Description} & {\bf $\mathcal{L}$ [fb$^{-1}$] } & {\bf UL} & {\bf EM}\\
\hline \hline
\href{https://atlas.web.cern.ch/Atlas/GROUPS/PHYSICS/PAPERS/SUSY-2013-02/}{ATLAS-SUSY-2013-02}~\cite{Aad:2014wea} & 0$\ell$ + 2--6 jets & 20.3 &   & \checkmark  \\
\href{https://atlas.web.cern.ch/Atlas/GROUPS/PHYSICS/PAPERS/SUSY-2013-04/}{ATLAS-SUSY-2013-04}~\cite{Aad:2013wta} & 0$\ell$ + 7--10 jets & 20.3 &   & \checkmark \\
\href{https://atlas.web.cern.ch/Atlas/GROUPS/PHYSICS/PAPERS/SUSY-2013-09/}{ATLAS-SUSY-2013-09}~\cite{Aad:2014pda} & 2 SS $\ell$ + 0--3 $b$-jets & 20.3 &   & \checkmark  \\
\href{https://atlas.web.cern.ch/Atlas/GROUPS/PHYSICS/PAPERS/SUSY-2013-11/}{ATLAS-SUSY-2013-11}~\cite{Aad:2014vma} & 2$\ell$ ($e,\mu$), EWK & 20.3 &   & \checkmark \\
\href{https://atlas.web.cern.ch/Atlas/GROUPS/PHYSICS/PAPERS/SUSY-2013-12/}{ATLAS-SUSY-2013-12}~\cite{Aad:2014nua} & 3$\ell$ ($e,\mu,\tau$), EWK & 20.3 &   & \checkmark   \\
\href{https://atlas.web.cern.ch/Atlas/GROUPS/PHYSICS/PAPERS/SUSY-2013-18/}{ATLAS-SUSY-2013-18}~\cite{Aad:2014lra} & jets + $\geq$  3 $b$-jets & 20.1 &   & \checkmark  \\
\hline \hline
\href{http://cms-results.web.cern.ch/cms-results/public-results/publications/EXO-13-006/index.html}{CMS-EXO-13-006}~\cite{Khachatryan:2015lla} & HSCP & 18.8 &   & \checkmark  \\
\href{https://twiki.cern.ch/twiki/bin/view/CMSPublic/PhysicsResultsSUS13016}{CMS-PAS-SUS-13-016}~\cite{CMS-PAS-SUS-13-016} & 2 OS $\ell$ + $\geq$  4 (2 $b$-)jets & 19.7 &   & \checkmark   \\
\href{https://twiki.cern.ch/twiki/bin/view/CMSPublic/PhysicsResultsSUS12024}{CMS-SUS-12-024}~\cite{Chatrchyan:2013wxa} & 0$\ell$ + $\geq$  3 (1 $b$-)jets & 19.4 &   & \checkmark   \\
\href{https://twiki.cern.ch/twiki/bin/view/CMSPublic/PhysicsResultsSUS12028}{CMS-SUS-12-028}~\cite{Chatrchyan:2013mys} & multi ($b$-)jets, $\alpha_T$ & 11.7 & \checkmark &    \\
\href{https://twiki.cern.ch/twiki/bin/view/CMSPublic/PhysicsResultsSUS13007}{CMS-SUS-13-007}~\cite{Chatrchyan:2013iqa} & 1$\ell$ + $\geq$  2 $b$-jets & 19.3 &   & \checkmark   \\
\href{https://twiki.cern.ch/twiki/bin/view/CMSPublic/PhysicsResultsSUS13012}{CMS-SUS-13-012}~\cite{Chatrchyan:2014lfa} & jets + $\not{\!\!H}_T$ & 19.5 & \checkmark & \checkmark   \\
\href{https://twiki.cern.ch/twiki/bin/view/CMSPublic/PhysicsResultsSUS14010}{CMS-SUS-14-010}~\cite{CMS:2014dpa} & $b$-jets + 4 $W$ & 19.5 & \checkmark &    \\
\hline
\end{tabular}\end{table}

\clearpage

\end{appendix}



\nolinenumbers

\clearpage
\begin{thebibliography}{100}
\providecommand{\url}[1]{\texttt{#1}}
\providecommand{\urlprefix}{URL }
\expandafter\ifx\csname urlstyle\endcsname\relax
  \providecommand{\doi}[1]{doi:\discretionary{}{}{}#1}\else
  \providecommand{\doi}{doi:\discretionary{}{}{}\begingroup \urlstyle{rm}\Url}\fi
\providecommand{\eprint}[2][]{\href{https://arxiv.org/abs/#2}{arXiv:#2}}

\bibitem{Kraml:2013mwa}
S.~Kraml, S.~Kulkarni, U.~Laa, A.~Lessa, W.~Magerl, D.~Proschofsky-Spindler and W.~Waltenberger,
\newblock \emph{{SModelS: a tool for interpreting simplified-model results from the LHC and its application to supersymmetry}},
\newblock Eur. Phys. J. C \textbf{74}, 2868 (2014),
\newblock \doi{10.1140/epjc/s10052-014-2868-5},
\newblock \eprint{1312.4175}.

\bibitem{Altakach:2023tsd}
M.~M. Altakach, S.~Kraml, A.~Lessa, S.~Narasimha, T.~Pascal, T.~Reymermier and W.~Waltenberger,
\newblock \emph{{Global LHC constraints on electroweak-inos with SModelS v2.3}},
\newblock SciPost Phys. \textbf{16}(4), 101 (2024),
\newblock \doi{10.21468/SciPostPhys.16.4.101},
\newblock \eprint{2312.16635}.

\bibitem{ATLAS:2024qmx}
G.~Aad \emph{et~al.},
\newblock \emph{{ATLAS Run 2 searches for electroweak production of supersymmetric particles interpreted within the pMSSM}},
\newblock JHEP \textbf{05}, 106 (2024),
\newblock \doi{10.1007/JHEP05(2024)106},
\newblock \eprint{2402.01392}.

\bibitem{MSSMWorkingGroup:1998fiq}
A.~Djouadi \emph{et~al.},
\newblock \emph{{The Minimal supersymmetric standard model: Group summary report}},
\newblock In \emph{{GDR (Groupement De Recherche) - Supersymetrie}} (1998), \eprint{hep-ph/9901246}.

\bibitem{ATLAS:scandata}
\url{https://atlas.web.cern.ch/Atlas/GROUPS/PHYSICS/PAPERS/SUSY-2020-15/inputs/ATLAS_EW_pMSSM_Run2.html}.

\bibitem{Ambrogi:2017lov}
F.~Ambrogi, S.~Kraml, S.~Kulkarni, U.~Laa, A.~Lessa and W.~Waltenberger,
\newblock \emph{{On the coverage of the pMSSM by simplified model results}},
\newblock Eur. Phys. J. C \textbf{78}(3), 215 (2018),
\newblock \doi{10.1140/epjc/s10052-018-5660-0},
\newblock \eprint{1707.09036}.

\bibitem{Baum:2023inl}
S.~Baum, M.~Carena, T.~Ou, D.~Rocha, N.~R. Shah and C.~E.~M. Wagner,
\newblock \emph{{Lighting up the LHC with Dark Matter}},
\newblock JHEP \textbf{11}, 037 (2023),
\newblock \doi{10.1007/JHEP11(2023)037},
\newblock \eprint{2303.01523}.

\bibitem{Skands:2003cj}
P.~Z. Skands, B.~Allanach, H.~Baer, C.~Balazs, G.~Belanger \emph{et~al.},
\newblock \emph{{SUSY Les Houches accord: Interfacing SUSY spectrum calculators, decay packages, and event generators}},
\newblock JHEP \textbf{0407}, 036 (2004),
\newblock \doi{10.1088/1126-6708/2004/07/036},
\newblock \eprint{hep-ph/0311123}.

\bibitem{Ibe:2012sx}
M.~Ibe, S.~Matsumoto and R.~Sato,
\newblock \emph{{Mass Splitting between Charged and Neutral Winos at Two-Loop Level}},
\newblock Phys. Lett. B \textbf{721}, 252 (2013),
\newblock \doi{10.1016/j.physletb.2013.03.015},
\newblock \eprint{1212.5989}.

\bibitem{Ibe:2023dcu}
M.~Ibe, Y.~Nakayama and S.~Shirai,
\newblock \emph{{Precise estimate of charged Higgsino/Wino decay rate}},
\newblock JHEP \textbf{03}, 012 (2024),
\newblock \doi{10.1007/JHEP03(2024)012},
\newblock \eprint{2312.08087}.

\bibitem{Porod:2003um}
W.~Porod,
\newblock \emph{{SPheno, a program for calculating supersymmetric spectra, SUSY particle decays and SUSY particle production at e+ e- colliders}},
\newblock Comput. Phys. Commun. \textbf{153}, 275 (2003),
\newblock \doi{10.1016/S0010-4655(03)00222-4},
\newblock \eprint{hep-ph/0301101}.

\bibitem{ATLAS:2023tkt}
G.~Aad \emph{et~al.},
\newblock \emph{{Combination of searches for invisible decays of the Higgs boson using 139~fb$^{-1}$ of proton-proton collision data at $\sqrt{s}=13$~TeV collected with the ATLAS experiment}},
\newblock Phys. Lett. B \textbf{842}, 137963 (2023),
\newblock \doi{10.1016/j.physletb.2023.137963},
\newblock \eprint{2301.10731}.

\bibitem{ATLAS:2019nkf}
G.~Aad \emph{et~al.},
\newblock \emph{{Combined measurements of Higgs boson production and decay using up to $80$ fb$^{-1}$ of proton-proton collision data at $\sqrt{s}=13$~TeV collected with the ATLAS experiment}},
\newblock Phys. Rev. D \textbf{101}(1), 012002 (2020),
\newblock \doi{10.1103/PhysRevD.101.012002},
\newblock \eprint{1909.02845}.

\bibitem{Fuks:2012qx}
B.~Fuks, M.~Klasen, D.~R. Lamprea and M.~Rothering,
\newblock \emph{{Gaugino production in proton-proton collisions at a center-of-mass energy of 8 TeV}},
\newblock JHEP \textbf{10}, 081 (2012),
\newblock \doi{10.1007/JHEP10(2012)081},
\newblock \eprint{1207.2159}.

\bibitem{Fuks:2013vua}
B.~Fuks, M.~Klasen, D.~R. Lamprea and M.~Rothering,
\newblock \emph{{Precision predictions for electroweak superpartner production at hadron colliders with Resummino}},
\newblock Eur. Phys. J. C \textbf{73}, 2480 (2013),
\newblock \doi{10.1140/epjc/s10052-013-2480-0},
\newblock \eprint{1304.0790}.

\bibitem{Fiaschi:2018hgm}
J.~Fiaschi and M.~Klasen,
\newblock \emph{{Neutralino-chargino pair production at NLO+NLL with resummation-improved parton density functions for LHC Run II}},
\newblock Phys. Rev. D \textbf{98}(5), 055014 (2018),
\newblock \doi{10.1103/PhysRevD.98.055014},
\newblock \eprint{1805.11322}.

\bibitem{Fiaschi:2023tkq}
J.~Fiaschi, B.~Fuks, M.~Klasen and A.~Neuwirth,
\newblock \emph{{Electroweak superpartner production at 13.6 TeV with Resummino}},
\newblock Eur. Phys. J. C \textbf{83}(8), 707 (2023),
\newblock \doi{10.1140/epjc/s10052-023-11888-y},
\newblock \eprint{2304.11915}.

\bibitem{Nadolsky:2008zw}
P.~M. Nadolsky, H.-L. Lai, Q.-H. Cao, J.~Huston, J.~Pumplin, D.~Stump, W.-K. Tung and C.~P. Yuan,
\newblock \emph{{Implications of CTEQ global analysis for collider observables}},
\newblock Phys. Rev. D \textbf{78}, 013004 (2008),
\newblock \doi{10.1103/PhysRevD.78.013004},
\newblock \eprint{0802.0007}.

\bibitem{ATLAS:2021yqv}
G.~Aad \emph{et~al.},
\newblock \emph{{Search for charginos and neutralinos in final states with two boosted hadronically decaying bosons and missing transverse momentum in $pp$ collisions at $\sqrt{s} = 13$~TeV with the ATLAS detector}},
\newblock Phys. Rev. D \textbf{104}(11), 112010 (2021),
\newblock \doi{10.1103/PhysRevD.104.112010},
\newblock \eprint{2108.07586}.

\bibitem{ATLAS:2020pgy}
G.~Aad \emph{et~al.},
\newblock \emph{{Search for direct production of electroweakinos in final states with one lepton, missing transverse momentum and a Higgs boson decaying into two $b$-jets in $pp$ collisions at $\sqrt{s}=13$ TeV with the ATLAS detector}},
\newblock Eur. Phys. J. C \textbf{80}(8), 691 (2020),
\newblock \doi{10.1140/epjc/s10052-020-8050-3},
\newblock \eprint{1909.09226}.

\bibitem{ATLAS:2019lff}
G.~Aad \emph{et~al.},
\newblock \emph{{Search for electroweak production of charginos and sleptons decaying into final states with two leptons and missing transverse momentum in $\sqrt{s}=13$~TeV $pp$ collisions using the ATLAS detector}},
\newblock Eur. Phys. J. C \textbf{80}(2), 123 (2020),
\newblock \doi{10.1140/epjc/s10052-019-7594-6},
\newblock \eprint{1908.08215}.

\bibitem{ATLAS:2022zwa}
G.~Aad \emph{et~al.},
\newblock \emph{{Searches for new phenomena in events with two leptons, jets, and missing transverse momentum in 139~fb$^{-1}$ of $\sqrt{s}=13$~TeV $pp$ collisions with the ATLAS detector}},
\newblock Eur. Phys. J. C \textbf{83}(6), 515 (2023),
\newblock \doi{10.1140/epjc/s10052-023-11434-w},
\newblock \eprint{2204.13072}.

\bibitem{ATLAS:2021moa}
G.~Aad \emph{et~al.},
\newblock \emph{{Search for chargino-neutralino pair production in final states with three leptons and missing transverse momentum in $\sqrt{s} = 13$~TeV $pp$ collisions with the ATLAS detector}},
\newblock Eur. Phys. J. C \textbf{81}(12), 1118 (2021),
\newblock \doi{10.1140/epjc/s10052-021-09749-7},
\newblock \eprint{2106.01676}.

\bibitem{ATLAS:2021yyr}
G.~Aad \emph{et~al.},
\newblock \emph{{Search for supersymmetry in events with four or more charged leptons in 139 fb$^{-1}$ of $\sqrt{s}= 13$~TeV pp collisions with the ATLAS detector}},
\newblock JHEP \textbf{07}, 167 (2021),
\newblock \doi{10.1007/JHEP07(2021)167},
\newblock \eprint{2103.11684}.

\bibitem{ATLAS:2018eui}
M.~Aaboud \emph{et~al.},
\newblock \emph{{Search for chargino-neutralino production using recursive jigsaw reconstruction in final states with two or three charged leptons in proton-proton collisions at $\sqrt{s}=13$~TeV with the ATLAS detector}},
\newblock Phys. Rev. D \textbf{98}(9), 092012 (2018),
\newblock \doi{10.1103/PhysRevD.98.092012},
\newblock \eprint{1806.02293}.

\bibitem{ATLAS:2019lng}
G.~Aad \emph{et~al.},
\newblock \emph{{Searches for electroweak production of supersymmetric particles with compressed mass spectra in $\sqrt{s}=13$ TeV $pp$ collisions with the ATLAS detector}},
\newblock Phys. Rev. D \textbf{101}(5), 052005 (2020),
\newblock \doi{10.1103/PhysRevD.101.052005},
\newblock \eprint{1911.12606}.

\bibitem{ATLAS:2022rme}
G.~Aad \emph{et~al.},
\newblock \emph{{Search for long-lived charginos based on a disappearing-track signature using 136 fb$^{-1}$ of $pp$ collisions at $\sqrt{s}=13$~TeV with the ATLAS detector}},
\newblock Eur. Phys. J. C \textbf{82}(7), 606 (2022),
\newblock \doi{10.1140/epjc/s10052-022-10489-5},
\newblock \eprint{2201.02472}.

\bibitem{ATLAS:2017oal}
M.~Aaboud \emph{et~al.},
\newblock \emph{{Search for long-lived charginos based on a disappearing-track signature in $pp$ collisions at $\sqrt{s}=13$~TeV with the ATLAS detector}},
\newblock JHEP \textbf{06}, 022 (2018),
\newblock \doi{10.1007/JHEP06(2018)022},
\newblock \eprint{1712.02118}.

\bibitem{CMS:2022sfi}
A.~Tumasyan \emph{et~al.},
\newblock \emph{{Search for electroweak production of charginos and neutralinos at $\sqrt{s}=13$~TeV in final states containing hadronic decays of WW, WZ, or WH and missing transverse momentum}},
\newblock Phys. Lett. B \textbf{842}, 137460 (2023),
\newblock \doi{10.1016/j.physletb.2022.137460},
\newblock \eprint{2205.09597}.

\bibitem{MahdiAltakach:2023bdn}
M.~M. Altakach, S.~Kraml, A.~Lessa, S.~Narasimha, T.~Pascal and W.~Waltenberger,
\newblock \emph{{SModelS v2.3: Enabling global likelihood analyses}},
\newblock SciPost Phys. \textbf{15}(5), 185 (2023),
\newblock \doi{10.21468/SciPostPhys.15.5.185},
\newblock \eprint{2306.17676}.

\bibitem{lhcsxwg}
{LHC SUSY Cross Section Working Group},
\newblock \url{https://twiki.cern.ch/twiki/bin/view/LHCPhysics/SUSYCrossSections}.

\bibitem{Beenakker:1996ch}
W.~Beenakker, R.~Hopker, M.~Spira and P.~M. Zerwas,
\newblock \emph{{Squark and gluino production at hadron colliders}},
\newblock Nucl. Phys. B \textbf{492}, 51 (1997),
\newblock \doi{10.1016/S0550-3213(97)80027-2},
\newblock \eprint{hep-ph/9610490}.

\bibitem{Beenakker:2011fu}
W.~Beenakker, S.~Brensing, M.~n. Kramer, A.~Kulesza, E.~Laenen, L.~Motyka and I.~Niessen,
\newblock \emph{{Squark and Gluino Hadroproduction}},
\newblock Int. J. Mod. Phys. A \textbf{26}, 2637 (2011),
\newblock \doi{10.1142/S0217751X11053560},
\newblock \eprint{1105.1110}.

\bibitem{Beenakker:2013mva}
W.~Beenakker, T.~Janssen, S.~Lepoeter, M.~Kr{\"a}mer, A.~Kulesza, E.~Laenen, I.~Niessen, S.~Thewes and T.~Van~Daal,
\newblock \emph{{Towards NNLL resummation: hard matching coefficients for squark and gluino hadroproduction}},
\newblock JHEP \textbf{10}, 120 (2013),
\newblock \doi{10.1007/JHEP10(2013)120},
\newblock \eprint{1304.6354}.

\bibitem{Beenakker:2014sma}
W.~Beenakker, C.~Borschensky, M.~Kr{\"a}mer, A.~Kulesza, E.~Laenen, V.~Theeuwes and S.~Thewes,
\newblock \emph{{NNLL resummation for squark and gluino production at the LHC}},
\newblock JHEP \textbf{12}, 023 (2014),
\newblock \doi{10.1007/JHEP12(2014)023},
\newblock \eprint{1404.3134}.

\bibitem{Beenakker:2015rna}
W.~Beenakker, C.~Borschensky, M.~Kr{\"a}mer, A.~Kulesza, E.~Laenen, S.~Marzani and J.~Rojo,
\newblock \emph{{NLO+NLL squark and gluino production cross-sections with threshold-improved parton distributions}},
\newblock Eur. Phys. J. C \textbf{76}(2), 53 (2016),
\newblock \doi{10.1140/epjc/s10052-016-3892-4},
\newblock \eprint{1510.00375}.

\bibitem{Beenakker:2016lwe}
W.~Beenakker, C.~Borschensky, M.~Kr{\"a}mer, A.~Kulesza and E.~Laenen,
\newblock \emph{{NNLL-fast: predictions for coloured supersymmetric particle production at the LHC with threshold and Coulomb resummation}},
\newblock JHEP \textbf{12}, 133 (2016),
\newblock \doi{10.1007/JHEP12(2016)133},
\newblock \eprint{1607.07741}.

\bibitem{Kulesza:2008jb}
A.~Kulesza and L.~Motyka,
\newblock \emph{{Threshold resummation for squark-antisquark and gluino-pair production at the LHC}},
\newblock Phys. Rev. Lett. \textbf{102}, 111802 (2009),
\newblock \doi{10.1103/PhysRevLett.102.111802},
\newblock \eprint{0807.2405}.

\bibitem{Kulesza:2009kq}
A.~Kulesza and L.~Motyka,
\newblock \emph{{Soft gluon resummation for the production of gluino-gluino and squark-antisquark pairs at the LHC}},
\newblock Phys. Rev. D \textbf{80}, 095004 (2009),
\newblock \doi{10.1103/PhysRevD.80.095004},
\newblock \eprint{0905.4749}.

\bibitem{Fuks:2016vdc}
B.~Fuks, M.~Klasen and M.~Rothering,
\newblock \emph{{Soft gluon resummation for associated gluino-gaugino production at the LHC}},
\newblock JHEP \textbf{07}, 053 (2016),
\newblock \doi{10.1007/JHEP07(2016)053},
\newblock \eprint{1604.01023}.

\bibitem{CMS:2021beq}
A.~M. Sirunyan \emph{et~al.},
\newblock \emph{{Search for top squark production in fully-hadronic final states in proton-proton collisions at $\sqrt{s} = 13$ TeV}},
\newblock Phys. Rev. D \textbf{104}(5), 052001 (2021),
\newblock \doi{10.1103/PhysRevD.104.052001},
\newblock \eprint{2103.01290}.

\bibitem{Aaboud:2017mpt}
M.~Aaboud \emph{et~al.},
\newblock \emph{{Search for long-lived charginos based on a disappearing-track signature in pp collisions at $ \sqrt{s}=13 $ TeV with the ATLAS detector}},
\newblock JHEP \textbf{06}, 022 (2018),
\newblock \doi{10.1007/JHEP06(2018)022},
\newblock \eprint{1712.02118}.

\bibitem{Sirunyan:2019ctn}
A.~M. Sirunyan \emph{et~al.},
\newblock \emph{{Search for supersymmetry in proton-proton collisions at 13 TeV in final states with jets and missing transverse momentum}},
\newblock JHEP \textbf{10}, 244 (2019),
\newblock \doi{10.1007/JHEP10(2019)244},
\newblock \eprint{1908.04722}.

\bibitem{CMS:2020atg}
A.~M. Sirunyan \emph{et~al.},
\newblock \emph{{Search for disappearing tracks in proton-proton collisions at $\sqrt{s} = 13$ TeV}},
\newblock Phys. Lett. B \textbf{806}, 135502 (2020),
\newblock \doi{10.1016/j.physletb.2020.135502},
\newblock \eprint{2004.05153}.

\bibitem{Aad:2019vnb}
G.~Aad \emph{et~al.},
\newblock \emph{{Search for electroweak production of charginos and sleptons decaying into final states with two leptons and missing transverse momentum in $\sqrt{s}=13$ TeV $pp$ collisions using the ATLAS detector}},
\newblock Eur. Phys. J. \textbf{C80}(2), 123 (2020),
\newblock \doi{10.1140/epjc/s10052-019-7594-6},
\newblock \eprint{1908.08215}.

\bibitem{Aad:2019vvf}
G.~Aad \emph{et~al.},
\newblock \emph{{Search for direct production of electroweakinos in final states with one lepton, missing transverse momentum and a Higgs boson decaying into two $b$-jets in $pp$ collisions at $\sqrt{s}=13$ TeV with the ATLAS detector}},
\newblock Eur. Phys. J. \textbf{C80}(8), 691 (2020),
\newblock \doi{10.1140/epjc/s10052-020-8050-3},
\newblock \eprint{1909.09226}.

\bibitem{Sirunyan:2017lae}
A.~M. Sirunyan \emph{et~al.},
\newblock \emph{{Search for electroweak production of charginos and neutralinos in multilepton final states in proton-proton collisions at $\sqrt{s}=$ 13 TeV}},
\newblock JHEP \textbf{03}, 166 (2018),
\newblock \doi{10.1007/JHEP03(2018)166},
\newblock \eprint{1709.05406}.

\bibitem{CMS:2021edw}
A.~Tumasyan \emph{et~al.},
\newblock \emph{{Search for supersymmetry in final states with two or three soft leptons and missing transverse momentum in proton-proton collisions at $ \sqrt{s} $ = 13 TeV}},
\newblock JHEP \textbf{04}, 091 (2022),
\newblock \doi{10.1007/JHEP04(2022)091},
\newblock \eprint{2111.06296}.

\bibitem{Aad:2014vma}
G.~Aad \emph{et~al.},
\newblock \emph{{Search for direct production of charginos, neutralinos and sleptons in final states with two leptons and missing transverse momentum in $pp$ collisions at $\sqrt{s} =$ 8 TeV with the ATLAS detector}},
\newblock JHEP \textbf{05}, 071 (2014),
\newblock \doi{10.1007/JHEP05(2014)071},
\newblock \eprint{1403.5294}.

\bibitem{Aad:2014nua}
G.~Aad \emph{et~al.},
\newblock \emph{{Search for direct production of charginos and neutralinos in events with three leptons and missing transverse momentum in $\sqrt{s} =$ 8 TeV $pp$ collisions with the ATLAS detector}},
\newblock JHEP \textbf{04}, 169 (2014),
\newblock \doi{10.1007/JHEP04(2014)169},
\newblock \eprint{1402.7029}.

\bibitem{Aaboud:2017vwy}
M.~Aaboud \emph{et~al.},
\newblock \emph{{Search for squarks and gluinos in final states with jets and missing transverse momentum using 36 fb$^{-1}$ of $\sqrt{s}=13$ TeV pp collision data with the ATLAS detector}},
\newblock Phys. Rev. \textbf{D97}(11), 112001 (2018),
\newblock \doi{10.1103/PhysRevD.97.112001},
\newblock \eprint{1712.02332}.

\bibitem{CMS:2018kag}
A.~M. Sirunyan \emph{et~al.},
\newblock \emph{{Search for new physics in events with two soft oppositely charged leptons and missing transverse momentum in proton-proton collisions at $\sqrt{s}=$ 13 TeV}},
\newblock Phys. Lett. B \textbf{782}, 440 (2018),
\newblock \doi{10.1016/j.physletb.2018.05.062},
\newblock \eprint{1801.01846}.

\bibitem{Aaboud:2019trc}
M.~Aaboud \emph{et~al.},
\newblock \emph{{Search for heavy charged long-lived particles in the ATLAS detector in 36.1 fb$^{-1}$ of proton-proton collision data at $\sqrt{s} = 13$ TeV}},
\newblock Phys. Rev. \textbf{D99}(9), 092007 (2019),
\newblock \doi{10.1103/PhysRevD.99.092007},
\newblock \eprint{1902.01636}.

\bibitem{Aaboud:2018jiw}
M.~Aaboud \emph{et~al.},
\newblock \emph{{Search for electroweak production of supersymmetric particles in final states with two or three leptons at $\sqrt{s}=13\,$TeV with the ATLAS detector}},
\newblock Eur. Phys. J. \textbf{C78}(12), 995 (2018),
\newblock \doi{10.1140/epjc/s10052-018-6423-7},
\newblock \eprint{1803.02762}.

\bibitem{CMS:2020bfa}
A.~M. Sirunyan \emph{et~al.},
\newblock \emph{{Search for supersymmetry in final states with two oppositely charged same-flavor leptons and missing transverse momentum in proton-proton collisions at $\sqrt{s} =$ 13 TeV}},
\newblock JHEP \textbf{04}, 123 (2021),
\newblock \doi{10.1007/JHEP04(2021)123},
\newblock \eprint{2012.08600}.

\bibitem{Aad:2014wea}
G.~Aad \emph{et~al.},
\newblock \emph{{Search for squarks and gluinos with the ATLAS detector in final states with jets and missing transverse momentum using $\sqrt{s}=8$ TeV proton--proton collision data}},
\newblock JHEP \textbf{09}, 176 (2014),
\newblock \doi{10.1007/JHEP09(2014)176},
\newblock \eprint{1405.7875}.

\bibitem{Chatrchyan:2014lfa}
S.~Chatrchyan \emph{et~al.},
\newblock \emph{{Search for new physics in the multijet and missing transverse momentum final state in proton-proton collisions at $\sqrt{s}$= 8 TeV}},
\newblock JHEP \textbf{06}, 055 (2014),
\newblock \doi{10.1007/JHEP06(2014)055},
\newblock \eprint{1402.4770}.

\bibitem{Aad:2019vvi}
G.~Aad \emph{et~al.},
\newblock \emph{{Search for chargino-neutralino production with mass splittings near the electroweak scale in three-lepton final states in $\sqrt {s}$=13 TeV $pp$ collisions with the ATLAS detector}},
\newblock Phys. Rev. \textbf{D101}(7), 072001 (2020),
\newblock \doi{10.1103/PhysRevD.101.072001},
\newblock \eprint{1912.08479}.

\bibitem{Sirunyan:2017cwe}
A.~M. Sirunyan \emph{et~al.},
\newblock \emph{{Search for supersymmetry in multijet events with missing transverse momentum in proton-proton collisions at 13 TeV}},
\newblock Phys. Rev. \textbf{D96}(3), 032003 (2017),
\newblock \doi{10.1103/PhysRevD.96.032003},
\newblock \eprint{1704.07781}.

\bibitem{Aaboud:2016zdn}
M.~Aaboud \emph{et~al.},
\newblock \emph{{Search for squarks and gluinos in final states with jets and missing transverse momentum at $\sqrt{s} =$ 13 TeV with the ATLAS detector}},
\newblock Eur. Phys. J. \textbf{C76}(7), 392 (2016),
\newblock \doi{10.1140/epjc/s10052-016-4184-8},
\newblock \eprint{1605.03814}.

\bibitem{ATLAS:2020qlk}
G.~Aad \emph{et~al.},
\newblock \emph{{Search for direct production of electroweakinos in final states with missing transverse momentum and a Higgs boson decaying into photons in pp collisions at $ \sqrt{s} $ = 13 TeV with the ATLAS detector}},
\newblock JHEP \textbf{10}, 005 (2020),
\newblock \doi{10.1007/JHEP10(2020)005},
\newblock \eprint{2004.10894}.

\bibitem{Sirunyan:2017kqq}
A.~M. Sirunyan \emph{et~al.},
\newblock \emph{{Search for new phenomena with the $M_{\mathrm {T2}}$ variable in the all-hadronic final state produced in proton--proton collisions at $\sqrt{s} = 13$ $\,\text {TeV}$}},
\newblock Eur. Phys. J. \textbf{C77}(10), 710 (2017),
\newblock \doi{10.1140/epjc/s10052-017-5267-x},
\newblock \eprint{1705.04650}.

\bibitem{CMS:2019zmd}
A.~M. Sirunyan \emph{et~al.},
\newblock \emph{{Search for supersymmetry in proton-proton collisions at 13 TeV in final states with jets and missing transverse momentum}},
\newblock JHEP \textbf{10}, 244 (2019),
\newblock \doi{10.1007/JHEP10(2019)244},
\newblock \eprint{1908.04722}.

\bibitem{Araz:2022vtr}
J.~Y. Araz, A.~Buckley, B.~Fuks, H.~Reyes-Gonzalez, W.~Waltenberger, S.~L. Williamson and J.~Yellen,
\newblock \emph{{Strength in numbers: Optimal and scalable combination of LHC new-physics searches}},
\newblock SciPost Phys. \textbf{14}(4), 077 (2023),
\newblock \doi{10.21468/SciPostPhys.14.4.077},
\newblock \eprint{2209.00025}.

\bibitem{CMS:2021cox}
A.~Tumasyan \emph{et~al.},
\newblock \emph{{Search for electroweak production of charginos and neutralinos in proton-proton collisions at $ \sqrt{s} $ = 13 TeV}},
\newblock JHEP \textbf{04}, 147 (2022),
\newblock \doi{10.1007/JHEP04(2022)147},
\newblock \eprint{2106.14246}.

\bibitem{judita}
J.~Mamuzic,
\newblock {talk at the pMSSM taskforce meeting of the LHC REI WG on 4 Dec 2025, \url{https://indico.cern.ch/event/1616848/}}.

\bibitem{Agin:2025vgn}
D.~Agin, B.~Fuks, M.~D. Goodsell and T.~Murphy,
\newblock \emph{{A joint explanation for the soft lepton and monojet LHC excesses in the wino-bino model}},
\newblock Eur. Phys. J. C \textbf{85}(10), 1145 (2025),
\newblock \doi{10.1140/epjc/s10052-025-14886-4},
\newblock \eprint{2506.21676}.

\bibitem{Agin:2023yoq}
D.~Agin, B.~Fuks, M.~D. Goodsell and T.~Murphy,
\newblock \emph{{Monojets reveal overlapping excesses for light compressed higgsinos}},
\newblock Phys. Lett. B \textbf{853}, 138597 (2024),
\newblock \doi{10.1016/j.physletb.2024.138597},
\newblock \eprint{2311.17149}.

\bibitem{Chakraborti:2024pdn}
M.~Chakraborti, S.~Heinemeyer and I.~Saha,
\newblock \emph{{Consistent excesses in the search for $\tilde\chi_2^0\tilde\chi_1^\pm$: wino/bino vs.\ Higgsino dark matter}},
\newblock Eur. Phys. J. C \textbf{84}(8), 812 (2024),
\newblock \doi{10.1140/epjc/s10052-024-13180-z},
\newblock \eprint{2403.14759}.

\bibitem{Agin:2024yfs}
D.~Agin, B.~Fuks, M.~D. Goodsell and T.~Murphy,
\newblock \emph{{Seeking a coherent explanation of LHC excesses for compressed spectra}},
\newblock Eur. Phys. J. C \textbf{84}(11), 1218 (2024),
\newblock \doi{10.1140/epjc/s10052-024-13594-9},
\newblock \eprint{2404.12423}.

\bibitem{Martin:2024pxx}
S.~P. Martin,
\newblock \emph{{Implications of purity constraints on light Higgsinos}},
\newblock Phys. Rev. D \textbf{109}(9), 095045 (2024),
\newblock \doi{10.1103/PhysRevD.109.095045},
\newblock \eprint{2403.19598}.

\bibitem{ATLAS:2021kxv}
G.~Aad \emph{et~al.},
\newblock \emph{{Search for new phenomena in events with an energetic jet and missing transverse momentum in $pp$ collisions at $\sqrt {s}$ =13 TeV with the ATLAS detector}},
\newblock Phys. Rev. D \textbf{103}(11), 112006 (2021),
\newblock \doi{10.1103/PhysRevD.103.112006},
\newblock \eprint{2102.10874}.

\bibitem{CMS:2021far}
A.~Tumasyan \emph{et~al.},
\newblock \emph{{Search for new particles in events with energetic jets and large missing transverse momentum in proton-proton collisions at $ \sqrt{s} $ = 13 TeV}},
\newblock JHEP \textbf{11}, 153 (2021),
\newblock \doi{10.1007/JHEP11(2021)153},
\newblock \eprint{2107.13021}.

\bibitem{zenodoEntry}
L.~Constantin, S.~Kraml, A.~Lessa, T.~Reymermier and W.~Waltenberger,
\newblock \emph{{Resources for ``On the coverage of electroweak-inos within the pMSSM with SModelS - a comparison with the ATLAS pMSSM study'' [Data set]}},
\newblock \doi{10.5281/zenodo.17949022} (2025).

\bibitem{hepdata.104458.v1}
{ATLAS Collaboration},
\newblock \emph{{Search for charginos and neutralinos in final states with two boosted hadronically decaying bosons and missing transverse momentum in $pp$ collisions at $\sqrt{s}=13$ TeV with the ATLAS detector (Version 1, 2021-09-06)}},
\newblock {HEPData (collection)},
\newblock \url{https://doi.org/10.17182/hepdata.104458.v1} (2021).

\bibitem{hepdata.104458.v3.r9}
{ATLAS Collaboration},
\newblock \emph{{``{\tt FullLikelihood\_forHEPData.tar}'' of ``Search for charginos and neutralinos in final states with two boosted hadronically decaying bosons and missing transverse momentum in $pp$ collisions at $\sqrt{s}=13$ TeV with the ATLAS detector'' (Version 3, updated 2023-11-13)}},
\newblock {HEPData (other)},
\newblock \url{https://doi.org/10.17182/hepdata.104458.v3/r9} (2023).

\bibitem{Heinrich:2021gyp}
L.~Heinrich, M.~Feickert, G.~Stark and K.~Cranmer,
\newblock \emph{{pyhf: pure-Python implementation of HistFactory statistical models}},
\newblock J. Open Source Softw. \textbf{6}(58), 2823 (2021),
\newblock \doi{10.21105/joss.02823}.

\bibitem{pyhf}
L.~Heinrich, M.~Feickert and G.~Stark,
\newblock \emph{{pyhf: v0.6.3}},
\newblock \doi{10.5281/zenodo.1169739},
\newblock Https://github.com/scikit-hep/pyhf/releases/tag/v0.6.3.

\bibitem{CMS:SL}
{CMS Collaboration},
\newblock \emph{{Simplified likelihood for the re-interpretation of public CMS results}},
\newblock Tech. Rep. CMS-NOTE-2017-001, CERN, Geneva,
\newblock {\url{https://cds.cern.ch/record/2242860}} (2017).

\bibitem{Buckley:2018vdr}
A.~Buckley, M.~Citron, S.~Fichet, S.~Kraml, W.~Waltenberger and N.~Wardle,
\newblock \emph{{The Simplified Likelihood Framework}},
\newblock JHEP \textbf{04}, 064 (2019),
\newblock \doi{10.1007/JHEP04(2019)064},
\newblock \eprint{1809.05548}.

\bibitem{Aaboud:2018doq}
M.~Aaboud \emph{et~al.},
\newblock \emph{{Search for photonic signatures of gauge-mediated supersymmetry in 13 TeV $pp$ collisions with the ATLAS detector}},
\newblock Phys. Rev. \textbf{D97}(9), 092006 (2018),
\newblock \doi{10.1103/PhysRevD.97.092006},
\newblock \eprint{1802.03158}.

\bibitem{Aaboud:2018sua}
M.~Aaboud \emph{et~al.},
\newblock \emph{{Search for chargino-neutralino production using recursive jigsaw reconstruction in final states with two or three charged leptons in proton-proton collisions at $\sqrt{s}=13$ TeV with the ATLAS detector}},
\newblock Phys. Rev. \textbf{D98}(9), 092012 (2018),
\newblock \doi{10.1103/PhysRevD.98.092012},
\newblock \eprint{1806.02293}.

\bibitem{ATLAS:2021twp}
G.~Aad \emph{et~al.},
\newblock \emph{{Search for squarks and gluinos in final states with one isolated lepton, jets, and missing transverse momentum at $\sqrt{s}=13$~ with the ATLAS detector}},
\newblock Eur. Phys. J. C \textbf{81}(7), 600 (2021),
\newblock \doi{10.1140/epjc/s10052-021-09344-w},
\newblock \eprint{2101.01629}.

\bibitem{ATLAS:2020syg}
G.~Aad \emph{et~al.},
\newblock \emph{{Search for squarks and gluinos in final states with jets and missing transverse momentum using 139 fb$^{-1}$ of $\sqrt{s}$ =13 TeV $pp$ collision data with the ATLAS detector}},
\newblock JHEP \textbf{02}, 143 (2021),
\newblock \doi{10.1007/JHEP02(2021)143},
\newblock \eprint{2010.14293}.

\bibitem{ATLAS:2022pib}
G.~Aad \emph{et~al.},
\newblock \emph{{Search for heavy, long-lived, charged particles with large ionisation energy loss in $pp$ collisions at $\sqrt{s} = 13~\text{TeV}$ using the ATLAS experiment and the full Run 2 dataset}},
\newblock JHEP \textbf{2306}, 158 (2023),
\newblock \doi{10.1007/JHEP06(2023)158},
\newblock \eprint{2205.06013}.

\bibitem{ATLAS:2022hbt}
G.~Aad \emph{et~al.},
\newblock \emph{{Search for direct pair production of sleptons and charginos decaying to two leptons and neutralinos with mass splittings near the W-boson mass in $ \sqrt{s} $ = 13 TeV pp collisions with the ATLAS detector}},
\newblock JHEP \textbf{06}, 031 (2023),
\newblock \doi{10.1007/JHEP06(2023)031},
\newblock \eprint{2209.13935}.

\bibitem{Khachatryan:2017rhw}
V.~Khachatryan \emph{et~al.},
\newblock \emph{{Search for supersymmetry in the all-hadronic final state using top quark tagging in pp collisions at $\sqrt s = 13$ TeV}},
\newblock Phys. Rev. \textbf{D96}(1), 012004 (2017),
\newblock \doi{10.1103/PhysRevD.96.012004},
\newblock \eprint{1701.01954}.

\bibitem{Sirunyan:2017pjw}
A.~M. Sirunyan \emph{et~al.},
\newblock \emph{{Search for supersymmetry in proton-proton collisions at 13 TeV using identified top quarks}},
\newblock Phys. Rev. D \textbf{97}(1), 012007 (2018),
\newblock \doi{10.1103/PhysRevD.97.012007},
\newblock \eprint{1710.11188}.

\bibitem{Sirunyan:2019hzr}
A.~M. Sirunyan \emph{et~al.},
\newblock \emph{{Search for supersymmetry in events with a photon, jets, $\mathrm {b}$ -jets, and missing transverse momentum in proton--proton collisions at 13 $\,\text {Te}\text {V}$}},
\newblock Eur. Phys. J. \textbf{C79}(5), 444 (2019),
\newblock \doi{10.1140/epjc/s10052-019-6926-x},
\newblock \eprint{1901.06726}.

\bibitem{CMS:2020cpy}
A.~M. Sirunyan \emph{et~al.},
\newblock \emph{{Search for physics beyond the standard model in events with jets and two same-sign or at least three charged leptons in proton-proton collisions at $\sqrt{s}=$ 13 TeV}},
\newblock Eur. Phys. J. C \textbf{80}(8), 752 (2020),
\newblock \doi{10.1140/epjc/s10052-020-8168-3},
\newblock \eprint{2001.10086}.

\bibitem{CMS:2022vpy}
A.~Tumasyan \emph{et~al.},
\newblock \emph{{Search for higgsinos decaying to two Higgs bosons and missing transverse momentum in proton-proton collisions at $ \sqrt{s} $ = 13 TeV}},
\newblock JHEP \textbf{05}, 014 (2022),
\newblock \doi{10.1007/JHEP05(2022)014},
\newblock \eprint{2201.04206}.

\bibitem{Aad:2013wta}
G.~Aad \emph{et~al.},
\newblock \emph{{Search for new phenomena in final states with large jet multiplicities and missing transverse momentum at $\sqrt{s}$=8 TeV proton-proton collisions using the ATLAS experiment}},
\newblock JHEP \textbf{10}, 130 (2013),
\newblock \doi{10.1007/JHEP10(2013)130, 10.1007/JHEP01(2014)109},
\newblock [Erratum: JHEP01,109(2014)],
\newblock \eprint{1308.1841}.

\bibitem{Aad:2014pda}
G.~Aad \emph{et~al.},
\newblock \emph{{Search for supersymmetry at $\sqrt{s}$=8 TeV in final states with jets and two same-sign leptons or three leptons with the ATLAS detector}},
\newblock JHEP \textbf{06}, 035 (2014),
\newblock \doi{10.1007/JHEP06(2014)035},
\newblock \eprint{1404.2500}.

\bibitem{Aad:2014lra}
G.~Aad \emph{et~al.},
\newblock \emph{{Search for strong production of supersymmetric particles in final states with missing transverse momentum and at least three $b$-jets at $\sqrt{s}$= 8 TeV proton-proton collisions with the ATLAS detector}},
\newblock JHEP \textbf{10}, 024 (2014),
\newblock \doi{10.1007/JHEP10(2014)024},
\newblock \eprint{1407.0600}.

\bibitem{Khachatryan:2015lla}
V.~Khachatryan \emph{et~al.},
\newblock \emph{{Constraints on the pMSSM, AMSB model and on other models from the search for long-lived charged particles in proton-proton collisions at sqrt(s) = 8 TeV}},
\newblock Eur. Phys. J. \textbf{C75}(7), 325 (2015),
\newblock \doi{10.1140/epjc/s10052-015-3533-3},
\newblock \eprint{1502.02522}.

\bibitem{CMS-PAS-SUS-13-016}
{CMS Collaboration},
\newblock \emph{{Search for supersymmetry in pp collisions at sqrt(s) = 8 TeV in events with two opposite sign leptons, large number of jets, b-tagged jets, and large missing transverse energy.}},
\newblock Tech. Rep. CMS-PAS-SUS-13-016, CERN, Geneva (2013).

\bibitem{Chatrchyan:2013wxa}
S.~Chatrchyan \emph{et~al.},
\newblock \emph{{Search for Gluino Mediated Bottom- and Top-Squark Production in Multijet Final States in $pp$ Collisions at 8 TeV}},
\newblock Phys. Lett. \textbf{B725}, 243 (2013),
\newblock \doi{10.1016/j.physletb.2013.06.058},
\newblock \eprint{1305.2390}.

\bibitem{Chatrchyan:2013mys}
S.~Chatrchyan \emph{et~al.},
\newblock \emph{{Search for Supersymmetry in Hadronic Final States with Missing Transverse Energy Using the Variables $\alpha_T$ and b-Quark Multiplicity in $pp$ collisions at $\sqrt{s}= 8$ TeV}},
\newblock Eur. Phys. J. \textbf{C73}(9), 2568 (2013),
\newblock \doi{10.1140/epjc/s10052-013-2568-6},
\newblock \eprint{1303.2985}.

\bibitem{Chatrchyan:2013iqa}
S.~Chatrchyan \emph{et~al.},
\newblock \emph{{Search for Supersymmetry in pp Collisions at $\sqrt{s}$=8 TeV in Events with a Single Lepton, Large Jet Multiplicity, and Multiple b Jets}},
\newblock Phys. Lett. \textbf{B733}, 328 (2014),
\newblock \doi{10.1016/j.physletb.2014.04.023},
\newblock \eprint{1311.4937}.

\bibitem{CMS:2014dpa}
V.~Khachatryan \emph{et~al.},
\newblock \emph{{Searches for supersymmetry based on events with b jets and four W bosons in pp collisions at 8 TeV}},
\newblock Phys. Lett. \textbf{B745}, 5 (2015),
\newblock \doi{10.1016/j.physletb.2015.04.002},
\newblock \eprint{1412.4109}.

\end{thebibliography}
\end{document}